\documentclass[12pt,preprint]{aastex}
\usepackage{graphicx}
\usepackage{natbib}
\usepackage{color}
\begin{document}
\title{Magneto-hydrodynamic Simulations of a Jet Drilling an HI Cloud: Shock Induced Formation of Molecular Clouds and Jet Breakup}
\author{Yuta \textsc{Asahina}$^{1}$, Takayuki \textsc{Ogawa}$^{1}$, Tomohisa \textsc{Kawashima}$^{2}$, Naoko \textsc{Furukawa}$^{3}$, Rei
\textsc{Enokiya}$^{3}$, Hiroaki \textsc{Yamamoto}$^{3}$, Yasuo \textsc{Fukui}$^{3}$ and Ryoji \textsc{Matsumoto}$^{1}$}

\email{asahina@astro.s.chiba-u.ac.jp}

\affil{$^{1}$Department of Physics, Graduate School of Science, Chiba
University, 1-33 Yayoi-cho, Inage-ku, Chiba 263-8522, Japan}
\affil{$^{2}$Key Laboratory for Research in Galaxies and Cosmology, Shanghai Astronomical Observatory, Chinese Academy of Science, 80 Nandan Road, Shanghai 200030, China}
\affil{$^{3}$Department of Physics, Nagoya University, Furo-cho, Chikusa-ku,
Nagoya 464-8602, Japan}
\email{ }

%\maketitle

\begin{abstract}
 The formation mechanism of the jet-aligned CO clouds found by NANTEN CO
 observations is studied by magnetohydrodynamical (MHD)
 simulations taking into account the cooling of the
 interstellar medium. 
 Motivated by the association of the CO 
 clouds with the enhancement of HI gas density, we carried out MHD
 simulations of the propagation of a supersonic jet injected into the
 dense HI gas. 
 We found that the HI gas compressed by the bow shock ahead of
 the jet is cooled down 
 by growth of the cooling instability triggered by the density
 enhancement.
 As a result, cold dense sheath is formed around the interface between
 the jet and the HI gas. 
 The radial speed of the cold, dense gas in the sheath is a few $\mathrm{km\ s^{-1}}$ almost independent of the jet speed.
 Molecular clouds can be formed in this region.
 Since the dense sheath wrapping  the jet reflects waves generated in the cocoon, the jet is strongly perturbed by the vortices of the warm gas in the cocoon, which breaks up the jet and forms a secondary shock in the HI-cavity drilled by the jet.  The particle acceleration at
 the shock can be the origin of radio and X-ray filaments observed near the eastern edge
 of W50 nebula surrounding the galactic jet source
 SS433.
\end{abstract}


\section{Introduction}
Detailed analysis of the NANTEN $^{12}$CO $(J=1-0)$ survey of galactic
molecular gas revealed that molecular clouds are aligned with the jet 
ejected from the galactic jet source SS433 \citep{2008PASJ...60..715Y}. 
They locate along the extension of the major axis of the asymmetric radio nebula W50 \citep[e.g.,][]{1998AJ....116.1842D} and bipolar X-ray jets \citep{1994PASJ...46L.109Y, 1997ApJ...483..868S}.
The association of the HI density enhancement with molecular clouds indicates that the interaction of the jet and the HI cloud triggered 
the formation of the jet-aligned molecular clouds \citep{2008PASJ...60..715Y}.

The interaction of astrophysical jets with the ambient interstellar medium (ISM) has
been studied extensively by hydrodynamic and magnetohydrodynamic (MHD) 
simulations.
By carrying out axisymmetric, two-dimensional (2D) hydrodynamic simulations, \citet{1982A&A...113...285N} revealed basic structures of a supersonic jet 
consisting of a bow shock, cocoon, and working surface.
Subsequently, axisymmetric 2D MHD simulations of jet propagation were carried out \citep[e.g.,][]{1986ApJ...311L..63C, 1989ApJ...344...89L, 1990A&A...229..378K, 1992PASJ...44...245T}. 
\citet{1993ApJ...403...164} reported the results of three-dimensional MHD 
simulations of
a jet interacting with a dense gas cloud. 
However, radiative cooling was not included in these simulations.

\citet{1989ApJ...337L..37B} carried out axisymmetric 2D hydrodynamical simulations of supersonic jets including radiative cooling and found that in the parameter range appropriate for protostellar jets (jet temperature is $10^{4}\ \mathrm{K}$ and jet number density is $20\ \mathrm{cm^{-3}}$), a dense, cold shell condenses out of the shocked gas at the head of the jet because shock compression increases the density, and cooling rate. \citet{1990ApJ...360..370B} showed that when the jet density is smaller than the ambient density, the leading edge of the jet closely follows the contours of the cocoon because radiative cooling decreases the thermal pressure of the shocked ambient gas behind the bow shock \citep[see also ][]{1993ApJ...413..198S}. 
\citet{1998ApJ...494L..79F} and \citet{2000ApJ...540..192S} studied the effects of cooling on the propagation of magnetized protostellar jets by axisymmetric 2D MHD simulations. 
More recently, \citet{2008A&A...488..429T} reported the results of 2D MHD simulations of protostellar jets using a non-equilibrium, multispecies cooling function. 

The SS433 jet differs from protostellar jets in its speed and density. 
The jet speed, $0.26c$ \citep[see review by][]{1984ARA&A..22..507M} is much faster than protostellar jets. 
The jet temperature exceeds $10^{8}\ \mathrm{K}$ near the central object but decreases to $10^{5}\ \mathrm{K}$ within $10^{12}\ \mathrm{cm}$ from the central object \citep{1991A&A...241..112B}. 
X-ray observations around the eastern edge of W50 \citep{2007A&A...463..611B} indicate that the temperature of the X-ray emitting plasma is $0.3\ \mathrm{keV}$.
The heating source of the region can be the internal shock of the jet, which partially dissipates the kinetic energy of the jet. 

\begin{figure}[!ht]
   \plotone{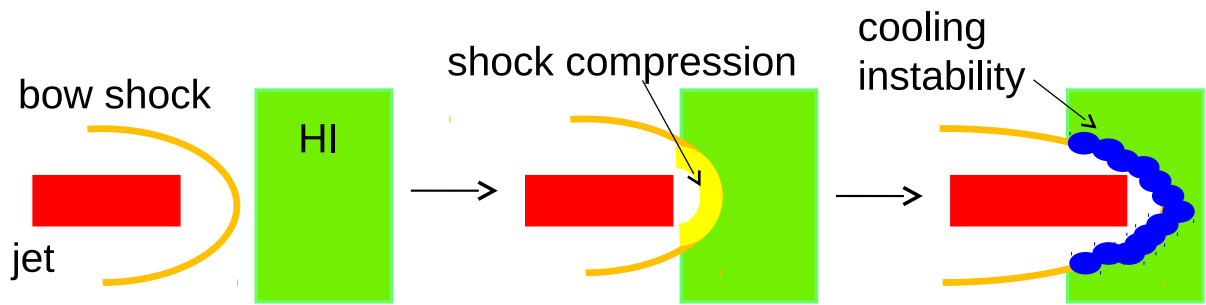}
  \caption{A schematic picture showing how cool dense clouds are formed
 in the interface between the supersonic jet and the HI cloud.  \label{coolpic}}
\end{figure}

Figure \ref{coolpic} schematically shows how the SS433 jet creates
molecular clouds.
When the jet propagates in the warm interstellar medium with number density $0.1\ \mathrm{cm^{-3}}$ and $T \sim 10^{4}\ \mathrm{K}$, cooling is negligible behind the bow shock (the left panel of Figure \ref{coolpic}).
However, when the supersonic jet is injected into the HI cloud, the density of the shock-compressed HI gas exceeds the threshold for the onset of the cooling instability (the right panel of Figure \ref{coolpic}). 
When the cooling instability is triggered and temperature decreases, the 
density increases further.
This mechanism is similar to that of formation of dense molecular gas by shock compression of the ISM \citep[e.g.,][]{2004ApJ...604...74F, 2006ApJ...652.1331I, 2009ApJ...695..825I}.
Molecular clouds can be formed around the surface of the HI-cavity drilled by the jet.

In this paper, we report the results of MHD simulations of jet propagation and interaction with an HI cloud 
including interstellar cooling. In section
2, we present numerical models. Numerical results are presented in
section 3. Section 4 is for summary and discussion.

\section{Numerical Model}

We carried out MHD simulations of jet propagation in a cylindrical
coordinate $(r,\phi,z)$.
We assume axisymmetry, but include the $\phi$-component of velocity and
magnetic field.
The basic equations of ideal MHD are
\begin{equation}
\frac{\partial \rho}{\partial t}+{\bf \nabla}
 \cdot \left (\rho \mbox{\boldmath $v$} \right) =0
\end{equation}
\begin{equation}
\frac{\partial \left( \rho \mbox{\boldmath $v$} \right)}{\partial t}
 +{\bf \nabla} \cdot \left(\rho \mbox{\boldmath $v$} \otimes
 \mbox{\boldmath $v$} +p+\frac{B^{2}}{8\pi} -\frac{\mbox{\boldmath $B$} \otimes
 \mbox{\boldmath $B$}}{4\pi}	\right)=0
\end{equation}
\begin{equation}
\frac{\partial}{\partial t} \left( e+\frac{B^{2}}{8\pi} \right)+{\bf \nabla} \cdot
 \left[\left( e+p \right) \mbox{\boldmath $v$}
 -\frac{ \left(\mbox{\boldmath $v$} \times \mbox{\boldmath $B$}\right) \times
 \mbox{\boldmath $B$}}{4\pi} \right]=-\rho L
\end{equation}
\begin{equation}
\frac{\partial \mbox{\boldmath $B$}}{\partial t}={\bf \nabla}
 \times \left(\mbox{\boldmath $v$} \times \mbox{\boldmath $B$}\right)
\end{equation}
where $\rho, \mbox{\boldmath $v$},p,\mbox{\boldmath $B$}$ are density, velocity,
pressure, and magnetic field, respectively, and
\begin{equation}
e=\frac{p}{\gamma -1}+\frac{\rho v^{2}}{2}
\end{equation}
is the energy density of the gas. In the right hand side of the energy equation (3),
$L$ is the cooling function.
We neglect thermal conduction, so that sharp temperature gradient can be preserved.
These equations are solved numerically by applying the HLLD scheme 
\citep{2005JComp...208...315}. 
HLLD scheme is a finite volume method to solve the magnetohydrodynamic equations in the conservation form, in which a Riemann problem at the cell interface 
is solved approximately by considering four intermediate states divided by two fast 
waves, two Alfv{\'e}n waves, and one entropy wave. 
HLLD scheme gives more accurate, less diffusive solutions than HLL scheme 
\citep{1983SIAM...25..35H} which takes into account only fast waves. 
Since the computational cost of the HLLD scheme is much less than the exact 
Riemann solver, and more easy to implement than other approximate Riemann solvers 
such as Roe scheme \citep{1981JCoPh..43..357R}, the HLLD scheme is widely used in magnetohydrodynamic 
simulations of astrophysical phenomena.
Second order accuracy in space is preserved by
linearly interpolating the values at the cell interface, and restricting them using the minmod limiter. 
To satisfy the solenoidal condition ${\bf \nabla} \cdot \mbox{\boldmath $B$}=0$, we
applied the generalized Lagrange multiplier (GLM) scheme proposed by \citet{2002JComp...175...645}. We incorporated the cooling term with time-implicit method.

\begin{figure}[!ht]
 \epsscale{0.9}
   \plotone{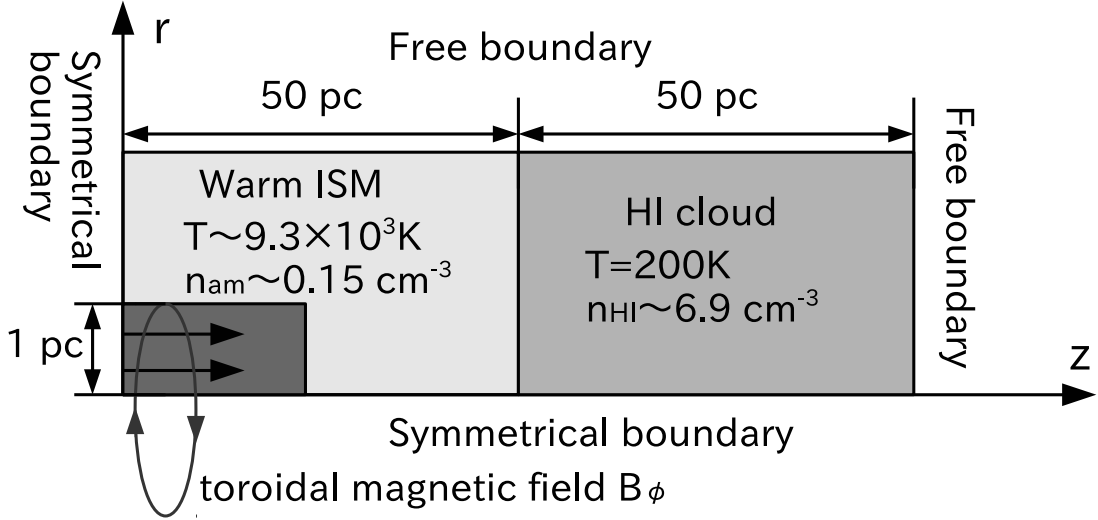}
  \caption{Simulation model. The simulation region is filled with the
 warm interstellar gas in $0 \leq z \leq 50\ \mathrm{pc}$ and HI gas in
 $50\ \mathrm{pc} \leq z \leq 100\ \mathrm{pc}$. Supersonic, hot jet is
 injected in $0 \leq r \leq 1\ \mathrm{pc}$, and at $z=0$. \label{model}}
\end{figure}

Figure \ref{model} shows the simulation model.
The simulation domain is $ 0 \leq r \leq 200\ \mathrm{pc}$ and $ 0
\leq z \leq 100\ \mathrm{pc}$.
We apply free boundary condition at $r=200\ \mathrm{pc}$ and $z=100\ \mathrm{pc}$ and
symmetrical boundary condition at $r=0$ and $z=0$. 
For adiabatic simulations, the number of grid points 
is $(N_{r},N_{z})=(500,1920)$. 
We used a uniform grid in $z$-direction.
In the radial direction, uniform grid with mesh size $0.054\ \mathrm{pc}$ is used in $0 \leq r \leq 15\ \mathrm{pc}$, so that the
jet radius ($r_{j}=1\ \mathrm{pc}$) is resolved with 19 cells. In $r \geq 15\ \mathrm{pc}$, we
increased the grid spacing with radius to avoid reflection at the outer radial boundary. For simulations
including the cooling, we used twice as many grid points 
, $(N_{r},N_{z}) = (900, 3820)$
to resolve the
thin, cold, dense region surrounding the jet.

At the initial state, the HI gas ($T \sim 200\ \mathrm{K}$) is assumed to be in pressure equilibrium with the warm
interstellar gas ($T \sim 10^{4}\ \mathrm{K}$) at $z = 50\ \mathrm{pc}$.
We assume that the HI gas and the warm interstellar gas are not magnetized at the initial state. 
The number density of the ambient medium and HI cloud are $n_{\mathrm{amb}}=0.15\ \mathrm{cm^{-3}}$, and $n_{\mathrm{HI}} \sim 6.9\ \mathrm{cm^{-3}}$, respectively. 
The initial density and temperature of the ISM are chosen such that they satisfy the thermal equilibrium condition $\rho L = 0$. 
We adopt the cooling function, 
\begin{equation}
\rho L = n(-\Gamma +n\Lambda) \exp \left\{- \left[ \mathrm{max} \left(
  \frac{T}{7000}-1,0\right) \right] ^{4}\right\}
\end{equation}
\begin{equation}
\Gamma = 2 \times 10^{-26}\ \mathrm{ergs\ s^{-1}}
\end{equation}
\begin{eqnarray}
\Lambda & = & 7.3\times 10^{-21} \exp \left( \frac{-118400}{T+1500} \right) \nonumber \\
  & &+7.9\times 10^{-27}\exp \left( \frac{-200}{T} \right)\ \mathrm{ergs\ cm^{3}\ s^{-1}}
\end{eqnarray}
where $\Gamma$ and $\Lambda$ are heating rate and cooling rate, respectively. 
They have the same form as those used by \cite{2006ApJ...652.1331I} but 
we modified it by cutting off the cooling in hot region where $T>14000\ \mathrm{K}$ and adjusting the last term in Equation (8) such that the equilibrium temperature is $200\ \mathrm{K}$ when $n=n_{\mathrm{HI}}$. 
We reduced the cooling rate when $T > 10^{4}\ \mathrm{K}$ so that the jet stays in hot ($T_{\mathrm{jet}} \sim 10^{5}\ \mathrm{K}$), low density ($n_{\mathrm{jet}} < 0.1\ \mathrm{cm^{-3}}$) state.

\begin{figure}[!ht]
  \epsscale{0.8}
   \plotone{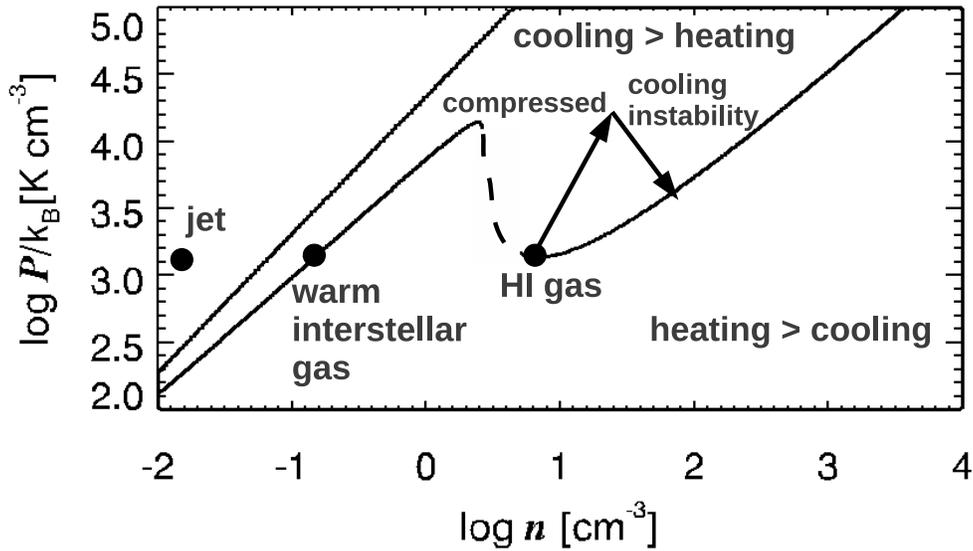}
  \caption{Thermal equilibrium curve for the cooling function adopted in this paper. The
  horizontal axis is the number density and the vertical axis is the
  pressure. Dashed curve indicates the unstable branch. \label{cooling}}
 \end{figure}

Figure \ref{cooling} shows the thermal equilibrium curves of the cooling function (
$\Gamma =n \Lambda$ in 
Equation (6)) adopted in this paper. The upper branch appears because we cut off the cooling for high temperature plasma. The lower branch has thermally stable branches (solid curve) and an unstable branch (dashed curve) connecting stable branches.

We inject a 
weakly 
magnetized jet from the injection region at $z = 0$ and $0 < r < r_{\mathrm{jet}} = 1\ \mathrm{pc}$. The number density and temperature of the injected plasma 
in a canonical model
are $n_{\mathrm{jet}} \sim 0.015\ \mathrm{cm^{-3}} = 0.1n_{\mathrm{amb}}$ and $T_{\mathrm{jet}} = 10^{5}\ \mathrm{K}$, respectively. 
The injection speed of the jet $v_{\mathrm{jet}}$ is chosen to be much smaller than that of the subrelativistic 
jet speed in SS433 ($v_{\mathrm{jet}} = 0.26c$)
because we have to deal with very high Mach number flows. 
Here the Mach number is defined as $v_{\mathrm{jet}}/c_{s\mathrm{,jet}}$, where 
$c_{s\mathrm{,jet}}$ is the sound speed in the jet. 
When we numerically solve the conservation form of energy equation to handle strong 
shocks, numerical accuracy of gas pressure degrades in high Mach number flows because gas pressure is derived by subtracting kinetic energy from the total energy. 
Therefore, we carry out simulations for flows with Mach number less than 20. 
Still, we can study the propagation of supersonic jet, in which the dynamical 
pressure $r_{\mathrm{jet}}v_{\mathrm{jet}}^{2}$ much exceeds the thermal pressure 
of ambient medium. 
We study the dependence of numerical results on $v_{\mathrm{jet}}$, and extrapolate 
the results to extremely high Mach number flows. 

Table 1 shows the model parameters. 
MA6 is a model without cooling. 
Other models are models with cooling.
For model MA6, MC6 and MC19H
 the mass flux of the jet $\dot M_{\mathrm{jet}} = \rho_{\mathrm{jet}} v_{\mathrm{jet}} \pi r_{\mathrm{jet}}^{2} \sim 10^{19}\ \mathrm{g\ s^{-1}}$ is chosen to be comparable to that of SS433 ($\dot M_{\mathrm{jet}} \sim 10^{19} \mathrm{g\ s^{-1}}$).
The jet speed for models MC3, MC6 and MC12 (Mach 3, 6 and 12) are $v_{\mathrm{jet}} = 110\ \mathrm{km\ s^{-1}}, 220\ \mathrm{km\ s^{-1}}$ and $440\ \mathrm{km\ s^{-1}} $.
MC6H and MC19H are models with higher jet temperature ($T_{\mathrm{jet}} \sim 10^{6}\ \mathrm{K}$) with cooling.
The jet speed for models MC6H and MC19H are $v_{\mathrm{jet}} \sim 680\ \mathrm{km\ s^{-1}}$ and $2200\ \mathrm{km\ s^{-1}}$, respectively.
For magnetic fields, we assume that the jet is injected with purely toroidal
magnetic field $B_{\phi}\propto \sin ^{4}(2\pi r/r_{\mathrm{jet}})$. We assumed $P_{\mathrm{gas}}/P_{\mathrm{mag}} = 100$ at $r = 0.5\ \mathrm{pc}$. 
In pointing flux dominated jets, it is possible that $P_{\mathrm{mag}}$ exceeds 
$P_{\mathrm{gas}}$. However, we assumed weakly magnetized jets to study how the magnetic 
fields are amplified during the propagation of the jet, and by cooling. 
The effects of strong toroidal magnetic fields on the stability of the jet will be 
studied in subsequent papers by three-dimensional magnetohydrodynamic simulations 
including cooling.

\section{Numerical Results}
\subsection{Formation of Cold, Dense Sheath in Simulations with Cooling}

\begin{figure*}[!ht]
 \epsscale{1.0}
   \plotone{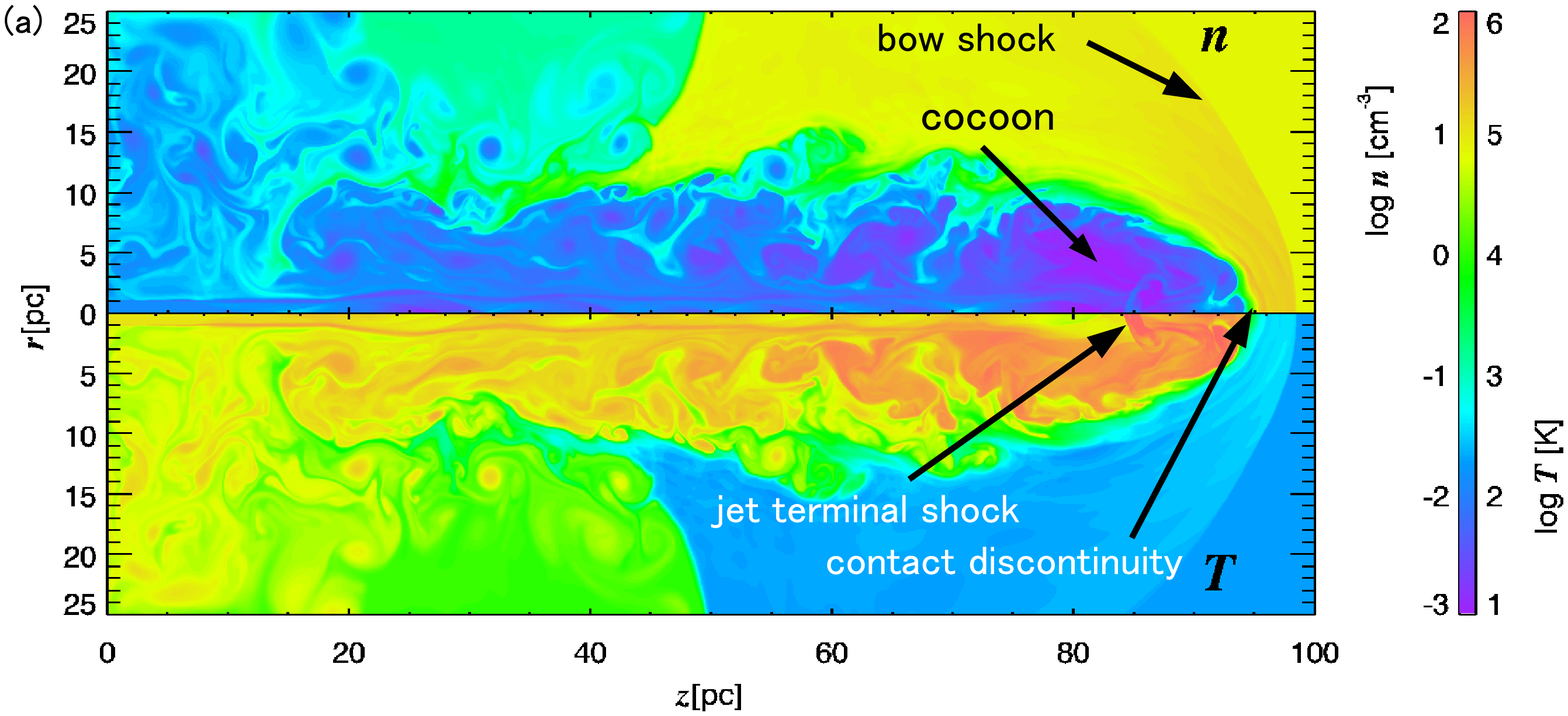} 
   \plotone{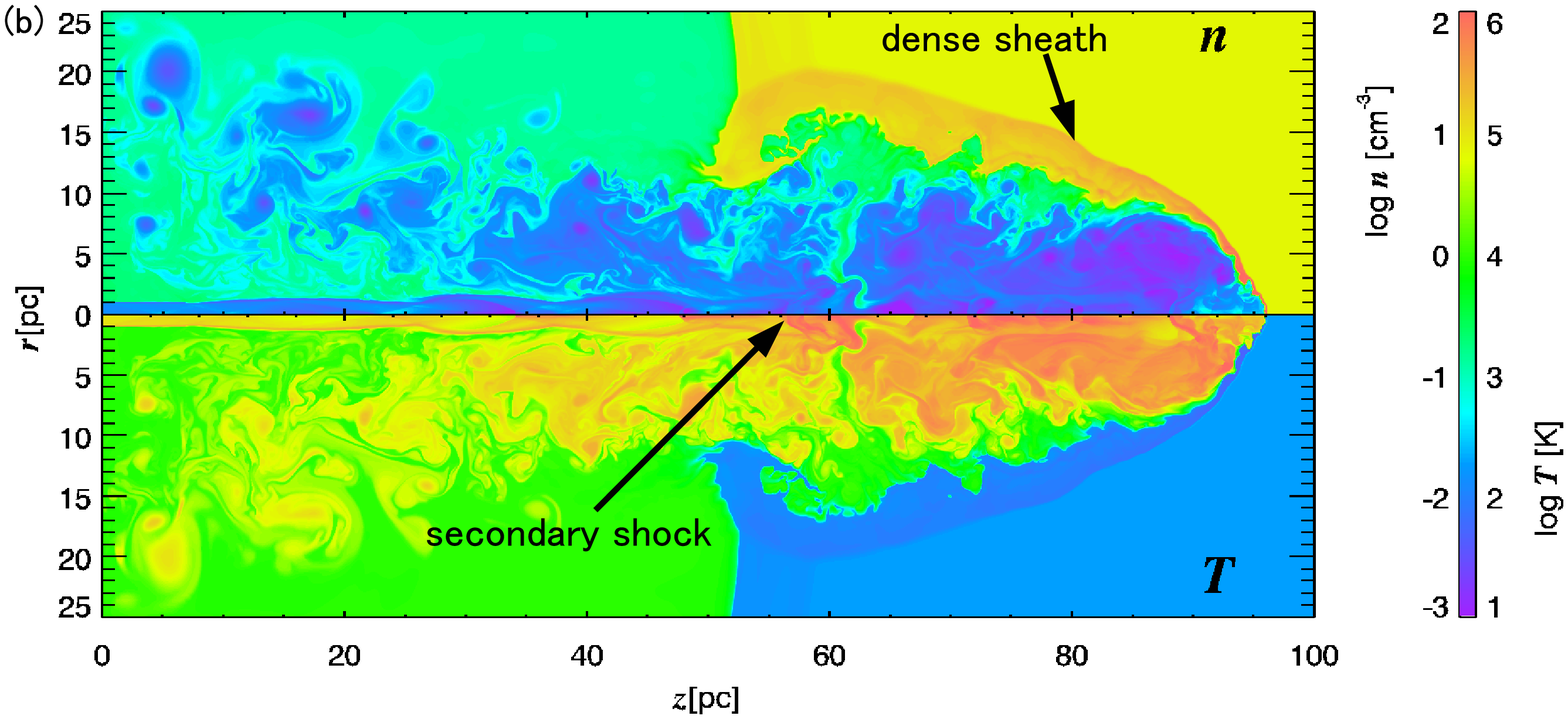} 
  \caption{The density and temperature distribution. Results for 
    (a) MA6, (b) MC6 at $t=14.7\ \mathrm{Myr}$. \label{MHDc}}
\end{figure*}

Figure \ref{MHDc} shows the results of numerical simulation without cooling (model MA6) and with cooling (model MC6). 
Color shows the distribution of density and temperature.
Figure \ref{MHDc} (a) shows results of an adiabatic MHD simulation.
Low density, hot jet ($T=10^{5}\ \mathrm{K}$) propagating in the warm interstellar gas collides with the HI cloud at $z=50\ \mathrm{pc}$. The jet
forms a bow shock, a jet terminal shock and internal shocks. 
The HI gas flowing in through the bow shock is compressed, and
heated up. 
The temperature and density of the HI gas increase to $T \sim 500\ \mathrm{K}$ and $n
\sim 20\ \mathrm{cm^{-3}}$, respectively. 
The jet gas between the jet terminal shock and the contact discontinuity
separating the jet and ambient medium, is 
heated up to $T \sim 10^{6}\ \mathrm{K}$ and the
velocity along the jet axis reverses, forming a hot, low density
backflow wrapping the jet (cocoon). The Kelvin-Helmholtz instability grows between the cocoon
and interstellar medium by their velocity shear. 
These structures are common to those in 
conventional
simulations \citep[e.g.,][]{1982A&A...113...285N}. 

Figure \ref{MHDc} (b) shows results of magnetohydrodynamical simulations including cooling.
Before collision, the jet structures are the same as those of adiabatic simulations since we cut off the cooling in the jet.
After collision the HI gas inflowing through the bow shock is compressed. 
Subsequently the HI gas is cooled down by
the cooling instability because cooling rate increases by enhanced
density, and forms cool, dense region behind the bow shock.
The bow shock and the shock compressed region in adiabatic simulations are converted to this cold dense sheath. 
As shown in Figure \ref{MHDc}, the jet is heated up to $T \sim 10^{6}\ \mathrm{K}$ by
the internal shock and the jet terminal shock.
In the cold sheath surrounding the jet, the density and temperature are about $ 30\ \mathrm{cm^{-3}}$ and $100\ \mathrm{K}$, respectively. 
This sheath is thinner in the jet head than in sides of the jet.
Molecular gas can be formed in these cold, dense region.

Numerical results indicate that secondary shock appears around $z=55\ \mathrm{pc}$ at $t=14.7\ \mathrm{Myr}$ in the model with cooling (see the bottom panel in figure 4). The shock appears more clearly in velocity distribution displayed in top panels of figure 5. Color shows the velocity component parallel to the jet axis, and arrows show velocity vectors. The jet corresponds to the beam in the region $r < 1\ \mathrm{pc}$ where $v_z=220\ \mathrm{km\ s^{-1}}$. The velocity reverses in the cocoon, and forms a backflow with speed $v_z \sim -50\ \mathrm{km\ s^{-1}}$. Black contours show the isocontours of the radial velocity in the dense sheath where $n > 10\ \mathrm{cm^{-3}}$. 
 We found that the jet is disrupted at $t=10.9\ \mathrm{Myr}$ around $z = 55\ \mathrm{pc}$. The disruption takes place because the dense sheath wrapping the cocoon reflects waves generated in the HI-cavity drilled by the jet. Since the vortices created in the cocoon are confined by the sheath, they strongly perturb the jet and block its propagation. The formation of the dense sheath results in the disruption of the jet. The top right panel of figure 5 shows that the jet beam in the HI-cavity is recovered at $t = 14.7\ \mathrm{Myr}$ but this beam is disconnected from the beam connecting the jet source and the initial surface of the HI cloud. 

The toroidal magnetic field
between the terminal shock and the bow shock becomes several times stronger than that before crossing the shock.
After the jet collides with the HI gas, the magnetic field is stored between
the cocoon and the sheath and becomes about 10 times stronger than
that injected into the simulation region at $z=0$ (the bottom panel in Figure 5).

Figure \ref{zvrvz} shows the distribution of the mean radial velocity $v_{r}$ and $v_{z}$  in the dense sheath ($n > 7 \mathrm{cm}^{-3}$) 
for model MC6. 
In the region around the head of the jet, the cold dense sheath moves along the jet axis with velocity $v_{z} \sim 3 \mathrm{km\ s^{-1}}$. 
This is the speed of the working surface of the jet, which can be computed by dividing the distance the jet propagates in the HI layer ($50\ \mathrm{pc}$) by the jet crossing time $14.7\ \mathrm{Myr}$. 
The radial velocity at the jet head, $v_{r} \sim 1.5 \mathrm{km\ s}^{-1}$ is about half of the speed of the working surface. 
On the other hand, in the sides of the jet, the sheath expands mainly in the radial direction with speed $v_{r} = 0.5-1\ \mathrm{km\ s^{-1}}$. 
The radial velocity is nearly constant when $z > 60\ \mathrm{pc}$ except the region near the head of the jet, where $v_{r} \sim 1.5\ \mathrm{km\ s}^{-1}$.

Figure \ref{sig} shows the column number density obtained by assuming that the HI cloud is a cylinder with radius $50\ \mathrm{pc}$. 
The column number density is about $10^{21.5}\ \mathrm{cm^{-2}}$ in the dense sheath formed around 
the jet-cloud interface. 
Figure \ref{sigh} shows the column number density of $\mathrm{H_{2}}$ when we assume solar abundance and neglect 
background UV radiation \citep[e.g.,][]{2014MNRAS.440.3349R}. 
The column number density $\sim 10^{21}\ \mathrm{cm^{-2}}$ in the dense sheath is consistent with 
the $\mathrm{H_{2}}$ column number density of molecular clouds aligned with the SS433 jet 
\citep[see Table I in ][]{2008PASJ...60..715Y}. 
We can obtain the CO intensity from the column number density of $\mathrm{H_{2}}$ by using the X factor for typical molecular clouds in the Galactic plane, 
$N(\mathrm{H_{2}})/W(\mathrm{^{12}CO})=2.0 \times 10^{20}\ \mathrm{cm^{-2}/(K\ km\ s^{-1})}$ 
\citep{1983ApJ...274..231L,1993ApJ...416..587B}.
The CO intensity $1-10\ \mathrm{K\ km\ s^{-1}}$ obtained from our numerical simulation is also consistent with the CO intensity $4-10\ \mathrm{K\ km\ s^{-1}}$ obtained by observations.

Figure \ref{jeth} shows the density and temperature distribution for 
the high temperature jet model 
MC6H, in which the temperature of the jet at the injection point is $T_{\mathrm{jet}}=10^{6}\ \mathrm{K}$. 
The propagation of the working surface of the jet and the shape of the dense sheath 
 are similar to those for model MC6, in which $T_{\mathrm{jet}}=10^{5}\ \mathrm{K}$. 
This is because the kinetic energy of the jet, $\rho _{\mathrm{jet}} v_{\mathrm{jet}}^{2}/2$, is the same for both models. 
It indicates that when the kinetic energy of the light jet ($\rho_{\mathrm{jet}} < \rho_{\mathrm{ambient}}$) is the same, 
the dynamics of the jet and the dense sheath
do not depend significantly on 
$v_{\mathrm{jet}}$ and 
the jet temperature (or the cooling function of the jet plasma).

\subsection{Dependence on the Beam Velocity}

In this subsection we show the dependence of numerical results on the beam velocity. 
Figure \ref{MHDc2} (a) shows results for a model with slower beam speed (MC3).
For MC3 
the speed of the jet working surface is $v_{\mathrm{ws}}=1.8\ \mathrm{km\ s}^{-1}$, which is about half of that for model MC6.
Since the velocity of the beam 
and the kinetic energy of the jet 
are smaller than that of MC6, the velocity of the backflow decreases. 
Since the cocoon becomes more turbulent when the backflow is slow, the beam tends to be disrupted, and does not extend to the jet head. The growth of the KH instability is observed in the region $z>60\ \mathrm{pc}$.
Similarly to model MC6, shock compression of the HI gas triggers the cooling instability which forms dense cold sheath surrounding the jet. 
The width of the sheath in the jet sides is larger than that of model MC6 because the jet propagation takes longer time, and the radial expansion speed of the dense sheath region is almost the same for both models.

Figure \ref{MHDc2} (b) shows results for 
Mach 12 jet (MC12). The speed of the jet working surface is $8\ \mathrm{km\ s^{-1}}$.
The faster backflow formed by the faster beam makes the cocoon more stable. 
The sheath around sides of the jet is thinner than other 
models because the propagation time of the jet is shorter. 

Let us compare the physical parameters of the SS433 jet with those of our model.
The jet speed $v_{\mathrm{jet}} \sim 220\ \mathrm{km\ s^{-1}}$ adopted in model MC6 is much slower than $0.26c$ measured by Doppler shifts of the line emission from the SS433 jet \citep{1989ApJ...347..448M}. 
We adopted smaller jet speed in our simulations to avoid numerical instabilities in high Mach number flows. 
The crossing time of the working surface ahead of the jet across the HI cloud with size $50\ \mathrm{pc}$ can be obtained by dividing this size with the speed of the working surface $v_{\mathrm{ws}}$. 
When the jet with speed $v_{\mathrm{jet}}=0.26c$ is injected into the HI cloud, the balance of the dynamical pressure of the jet and the ambient medium with density $\rho_{\mathrm{a}}$ gives \citep[e.g.,][]{1992PASJ...44...245T}
\begin{equation}
\rho_{\mathrm{jet}} r_{\mathrm{jet}}^{2} v_{\mathrm{jet}}^{2} = \rho_{\mathrm{a}} r_{\mathrm{ws}}^{2} v_{\mathrm{ws}}^{2}
\end{equation}
where $\rho_{\mathrm{jet}}$ is jet density and $r_{\mathrm{jet}}$ and $r_{\mathrm{ws}}$ are the width of the jet and working surface respectively. 
Here we assumed $v_{\mathrm{jet}} \gg v_{\mathrm{ws}}$.
By using the mass flux of the jet $\dot{M}_{\mathrm{jet}}= \pi r_{\mathrm{jet}}^{2} \rho_{\mathrm{jet}} v_{\mathrm{jet}}$ and combining this equation with (9), we obtain
\begin{equation}
v_{\mathrm{ws}}=\sqrt{\frac{v_{\mathrm{jet}} \dot{M}_{\mathrm{jet}}}{\pi \rho_{\mathrm{a}} r_{\mathrm{ws}} }}
\end{equation}
When $v_{\mathrm{jet}} = 0.26\mathrm{c}$, $\dot{M}_{\mathrm{jet}}=10^{19}\ \mathrm{g\ s^{-1}}$ \citep{2002ApJ...564..941M}, $\rho_{\mathrm{a}}=3 \times 10^{-25}\ \mathrm{g\ cm^{-3}}$ and $r_{\mathrm{ws}}=4\ \mathrm{pc}$, we obtain $v_{\mathrm{ws}}=50\ \mathrm{km\ s^{-1}}$. 
This speed is 20 times faster than the speed of the working surface in our simulations (model MC6) because $v_{\mathrm{jet}}=0.26c$ is 400 times faster than $v_{\mathrm{jet}}=220\ \mathrm{km\ s^{-1}}$ in model MC6. The crossing time of the working surface across the HI cloud is $10^{6}\ \mathrm{yr}$ when $v_{\mathrm{ws}}=50\ \mathrm{km\ s^{-1}}$.

Figure \ref{zvr} shows the mean radial velocity for 
MC3, MC6 and MC12 measured at different time, $t=27.1, 14.7$ and $5.84\ \mathrm{Myr}$, respectively. 
The radial velocity at the jet head is about half of the speed of the working surface of the jet. 
The radial expansion speed of the dense sheath increases with the jet speed. 

Figure \ref{tvr} shows the time evolution of the mean radial velocity at $z=65\ \mathrm{pc}$. 

We can estimate the radial velocity 
of the dense gas at the sheath from the following argument. 
We assume the shape of the sheath to be a hollow cylinder whose inner and outer radius, density and radial velocity at $t=t_{0}$ are $r_{\mathrm{i}}, r_{\mathrm{o}}, \rho$ and $v_{r}$, respectively. 
We denote those at $t=t'$ as
$r_{\mathrm{i}}', r_{\mathrm{o}}', \rho'$ and $v_{r}'$. Mass conservation and momentum conservation equations are 
\begin{equation}
\rho \pi (r_{\mathrm{o}}^{2}-r_{\mathrm{i}}^{2})+\rho_{\mathrm{HI}} \pi (r_{\mathrm{o}}'^{2}-r_{\mathrm{o}}^{2})=\rho' \pi (r_{\mathrm{o}}'^{2}-r_{\mathrm{i}}'^{2})
\end{equation}
\begin{equation}
\rho \pi (r_{\mathrm{o}}^{2}-r_{\mathrm{i}}^{2})v_{r}=\rho' \pi (r_{\mathrm{o}}'^{2}-r_{\mathrm{i}}'^{2})v_{r}'
\end{equation}
Solving for $v_{r}'$, we obtain
\begin{equation}
v_{r}' = \left[ 1+\frac{\rho_{\mathrm{HI}} \left( r_{\mathrm{o}}'^{2}- r_{\mathrm{o}}^{2} \right)}{\rho \left(r_{\mathrm{o}}^{2}- r_{\mathrm{i}}^{2} \right)} \right]^{-1} v_{r}
\end{equation}
Substituting $v_{r}'=\mathrm{d}r_{\mathrm{o}}'/\mathrm{d}t$ into equation (13) and integrating it, we obtain
\begin{equation}
\frac{A}{3r_{\mathrm{o}}^{2}}r_{\mathrm{o}}'^{3}+(1-A)r_{\mathrm{o}}'=v_{r}t+(1-\frac{2}{3}A)r_{\mathrm{o}}
\end{equation}
where
\begin{equation}
A=\left[ \frac{\rho}{\rho_{\mathrm{HI}}} \left( 1-\frac{r_{\mathrm{i}}^{2}}{r_{\mathrm{o}}^{2}}  \right)  \right]^{-1}
\end{equation}
If the HI gas 
pushed radially outward by the jet expands only in
the radial direction, mass conservation equation is
\begin{equation}
\rho_{\mathrm{HI}}\pi r_{\mathrm{o}}^{2} = \rho \pi (r_{\mathrm{o}}^{2}-r_{\mathrm{i}}^{2})
\end{equation}
Substituting equation (16) into equation (15), we obtain $A=1$. 
Therefore, from equation (14), 
\begin{equation}
r_{\mathrm{o}}'=\left(3r_{\mathrm{o}}^{2}v_{r}t+r_{\mathrm{o}}^{3}\right)^{\frac{1}{3}} \label{rod}
\end{equation}
and
\begin{equation}
v_{r}'=\frac{\mathrm{d}r_{\mathrm{o}}'}{\mathrm{d}t}=r_{\mathrm{o}}^{2}v_{r}\left(3r_{\mathrm{o}}^{2}v_{r}t+r_{\mathrm{o}}^{3}\right)^{-\frac{2}{3}}
\end{equation}
Figure \ref{vrest} shows the time evolution 
of $v_{r}'$ for initial radial velocities 
$v_{r}=2, 20, 200\ \mathrm{km\ s^{-1}}$.
The radial velocity becomes a few $\mathrm{km\ s^{-1}}$ at $t \sim 10\ \mathrm{Myr}$ even if it's initially $200\ \mathrm{km\ s^{-1}}$. 
This radial velocity is consistent with the observed velocity $2-5\ \mathrm{km\ s^{-1}}$ derived from the line width $\Delta V$ of the composite spectrum of 
the molecular clouds aligned with the SS433 jet and $V_{LSR}$ reported in 
Table 1 in \cite{2008PASJ...60..715Y}.

For all models, the shape of the interface between the cocoon and the dense sheath is similar. 
We can determine its shape using equation (\ref{rod}) as
\begin{equation}
r=\left[3r_{\mathrm{i}}^{2}v_{r} \left(t-t_{0} \right) \right]^{\frac{1}{3}}
\end{equation}
where we rewrite $r_{\mathrm{o}}'$ to $r$ and $r_{\mathrm{o}}$ to $r_{\mathrm{i}}$ and ignore the second term in equation (\ref{rod}). 
Here, $t_{0}$ is the time when the jet head passes. The position of the jet head can be evaluated by using equation (9) as 
\begin{equation}
z=v_{\mathrm{ws}}t_{0}=Bv_{\mathrm{jet}}t_{0}
\end{equation}
\begin{equation}
B=\sqrt{\frac{\rho_{\mathrm{jet}}}{\rho_{\mathrm{HI}}}} \frac{r_{\mathrm{jet}}}{r_{\mathrm{ws}}}
\end{equation}
Eliminating $t_{0}$ from equation (20) and equation(21), the radius of the interface can be obtained as a function of $r, z, t$ as 
\begin{equation}
r^{3} = 3r_{\mathrm{i}}^{2} v_{r} \left( t-\frac{z}{Bv_{\mathrm{jet}}} \right)
\end{equation}
Assuming $v_{r}=Cv_{\mathrm{jet}}$ ($C$ is constant), 
\begin{equation}
r^{3} = 3r_{\mathrm{i}}^{2}C \left(v_{\mathrm{jet}}t-\frac{z}{B} \right)
\end{equation}
When $v_{\mathrm{jet}}t$ is the same, the shape of the interface is the same even if $v_{\mathrm{jet}}$ is different. 
In Figure \ref{intf}, color shows the result for MC6 at $t=14.7\ \mathrm{Myr}$ and black curve shows the interface determined from equation (23) when $r_{\mathrm{i}}=7.4\ \mathrm{pc}$, $r_{\mathrm{ws}}=3.15\ \mathrm{pc}$, $C=10^{-2}$ and $t=13.0\ \mathrm{Myr}$. 
Here $t=13.0\ \mathrm{Myr}$ is the time since the jet head collides with the HI cloud. 

Figure \ref{jtmass} shows the time evolution of the total mass of the cold, dense sheath. 
When the beam velocity is high, the total mass increases since the region where the number density is larger than $20\ \mathrm{cm^{-3}}$ increases. 
The total mass of the HI gas swept by the bow shock limit the maximum total mass. 
When the swept up HI gas is originally located in a cylinder whose radius and length are $r_{\mathrm{HI}}\ \mathrm{pc}$ and $h_{\mathrm{HI}}\ \mathrm{pc}$, respectively, we can estimate the maximum total mass 
\begin{equation}
M_{\mathrm{max}} \sim 0.47 M_{\odot}{r_{\mathrm{HI}}}^{2} h_{\mathrm{HI}} \left(\frac{n_{\mathrm{HI}}}{6.9\ \mathrm{cm^{-3}}} \right)
\end{equation}
where $M_{\odot}$ is the solar mass. 
For MC12, substituting $r_{\mathrm{HI}}=13\ \mathrm{pc}$ and $h_{\mathrm{HI}}=50\ \mathrm{pc}$ to equation (16), 
we get $M_{\mathrm{max}} \sim 4000 M_{\odot}$. 
Figure \ref{jtmass} shows that the total mass of the sheath is about $3200 M_{\odot}$ which is close to $M_{\mathrm{max}}$. 
Further increase of the beam velocity does not affect the total mass of the cold, dense gas.

Figure \ref{jetdmj} shows the result for a high Mach number model for which the mass flux of the jet is equal to that for model MC6. 
Since the kinetic energy of the jet is higher than that for model MC6, the cold dense sheath becomes thinner.
This result is more similar to that for model MC12 than that for model MC6 since the kinetic energy of the jet is closer to that for model MC12.

\section{Summary and Discussion}
We have shown by performing MHD simulations including interstellar cooling that
cold dense sheath surrounding the jet is formed when the low-density, supersonic jet collides with the cool ($T \sim 200\ \mathrm{K}$) HI cloud. 
The interaction of the jet with the HI cloud is essential for the transition of the shock compressed gas to the cold ($T \sim 10\ \mathrm{K}$) state. 
The density and temperature of the cold, dense gas is comparable to
those in molecular clouds. 
This mechanism can explain the origin of molecular clouds aligned with the X-ray jet of SS433 and their association with the HI cloud \citep{2008PASJ...60..715Y}. 
On the other hand, when the ambient medium is warm ISM with temperature $T \sim 10^{4}\ \mathrm{K}$, the shock compressed ISM stays in the warm state with higher density.

Observations indicate that the integrated intensity of HI cloud along the axis of the SS433 jet is about $800\ \mathrm{K\ km\
s^{-1}}$ \citep{2008PASJ...60..715Y}. 
The column number density estimated by using the conventional
factor $1.8 \times 10^{18}\ \mathrm{cm^{-2}/(K\ km\ s^{-1})}$ 
is $1.4 \times 10^{21}\ \mathrm{cm^{-2}}$. 
When the depth of the HI cloud is $50\ \mathrm{pc}$, we can estimate the number density of the HI cloud to be $10\ \mathrm{cm^{-3}}$, which is comparable to that we assumed in our simulation.

The present numerical simulations have shown that the formation of the molecular clouds by the jet requires a timescale of $10^{6}\ \mathrm{yrs}$. 
This is much longer than the age of W50, $2 \times 10^{4}\ \mathrm{yrs}$, estimated by assuming that W50 is a supernova remnant \citep{1980A&A....84..237G}. \cite{2011MNRAS.414.2838G} carried out hydrodynamical simulations of the SS433 jet and showed that the W50 radio shell is consistent with such a short timescale. 
We should, however, note an alternative mechanism that a radio shell like W50 is being formed by the winds from a supercritically accreting black hole with a mass accretion rate exceeding the Eddington limit. 
According to the radiation hydrodynamical simulations of supercritical black hole accretion \citep[e.g.,][]{2009PASJ...61..769K}, radiation pressure driven wind emanating from the accretion disk can inject energy of order $10^{38}\ \mathrm{erg\ s^{-1}}$ for life time of the supercritical accretion. 
If the life time of SS433 is $10^{6}\ \mathrm{yrs}$, the energy supplied by the wind ($\sim 10^{51}\ \mathrm{erg}$) is comparable to that by supernova explosion. 
Recently, asymmetric radio and X-ray bubbles similar to W50 are found around a microquasar S26 in NGC7793 \citep{2010Natur.466..209P,2010MNRAS.409..541S} and an ultraluminous X-ray source (ULX) IC 342 X-1 \citep{2012ApJ...749...17C}. 
They can be inflated by jets and outflows from a supercritically accreting black hole.  
Therefore, it is a viable alternative that W50 is a bubble which has been continuously driven by the winds from SS433 over the last $10^{6}\ \mathrm{yrs}$. 

X-ray observations \citep{1997ApJ...483..868S, 1996A&A...312..306B} and radio observations \citep{1981A&A....97..296D, 1987AJ.....94.1633E} of W50 reported that X-ray and radio are strong near the eastern edge of W50 where the SS433 jet is drilling the HI cloud. 
Our numerical results indicate that the radio filament is not necessarily be the jet terminal shock at the leading edge of the jet but a secondary shock formed in the HI-cavity (see the top left panel in Figure 5). 
The jet terminal shocks and jet internal shocks can
produce synchrotron emitting relativistic electrons.
In high Mach number jets such as MC19H, since the
Mach number at the jet terminal shock and the jet
internal shock can exceed 20, relativistic electrons
can be produced quickly by shock surfing mechanism
with the Buneman instability \citep{2013PhRvL.111u5003M}. The relativistic electrons will be
accelerated further by diffusive shock acceleration.
Since the jet terminal shock near the leading edge of the jet is recovered from time to time, faint radio and X-ray emission may be detected in the HI-cavity if this region is observed with high sensitivity. 

Let us discuss the effects of precession of the SS433 jet.
The current precession period and the precession angle
of the SS433 jet are 162 days and 20 degrees, respectively.
According to the three-dimensional relativistic hydrodynamic
simulations of the precessing jet \citep{2014A&A...561A..30M},
helical jet beam behaves like a piston, which creates a
bow shock in front of the jet head, and the lower-density
cocoon surrounding the beam. In their simulation, the width
of the helical beam and the cocoon are $0.1\ \mathrm{pc}$, and the size
of the simulation region is ($0.2\ \mathrm{pc}$, $0.1\ \mathrm{pc}$, $0.1\ \mathrm{pc}$).
Therefore, the whole precessing jet is contained in the
jet injection region in our simulations shown in figure 2.
In our simulations, instead of resolving the precessing
jet, which requires the time resolution of order 10 days,
and spacial resolution of order $0.001\ \mathrm{pc}$, we replaced
it with the wider, lower density jet. Furthermore, the
current precession angle (20 degrees) is inconsistent
with the elongated shape of the W50 nebulae, indicating
that the precession angle had to be smaller than the
current angle when W50 was formed \citep{2011MNRAS.414.2838G}.

In this paper, we neglected the cooling in the
jet where $n \sim 0.01\ \mathrm{cm^{-3}}$ and $T \sim 10^{5}\ \mathrm{K}$.
If we take into account the cooling in the region
where $T > 10^{4}\ \mathrm{K}$, the cooling time scale of the jet
is the order of $0.1\ \mathrm{Myr}$ \citep[e.g.,][]{1993ApJS...88..253S},
Since the jet with $v_{\mathrm{jet}} \sim 200\ \mathrm{km\ s^{-1}}$ propagates about
$20\ \mathrm{pc}$ in $0.1\ \mathrm{Myr}$, the internal plasma of the jet can be
heated up by internal shocks of the jet within this
time scale. In our simulations in which the jet cooling
is neglected, the temperature of the jet exceeds the
original temperature ($T_{\mathrm{jet}} \sim 10^{5}\ \mathrm{K}$). We expect that
even when the cooling in the hot plasma is taken into
account, the jet temperature will stay $10^{5}\ \mathrm{K}  < T < 10^{6}\ \mathrm{K}$.
In subsequent papers, we would like to confirm it
by carrying out simulations including the cooling when
$T>10^{4}\ \mathrm{K}$ and thermal conduction, which affects the thermal
balance of the hot plasma.

In the jet head between the jet terminal shock and
the contact discontinuity and in the backflow region
(cocoon), the plasma temperature exceeds $10^{6}\ \mathrm{K}$ when
the jet speed exceeds $200\ \mathrm{km\ s^{-1}}$. Since the cooling time
scale of this region exceeds $10\ \mathrm{Myr}$, cooling can be
neglected.

The cooling time scale of the region between the
contact discontinuity and the bow shock where
$n > 10\ \mathrm{cm^{-3}}$ is the order of $10^{4}\ \mathrm{yr}$. Therefore,
this region cools down, and forms dense, cold
sheath.

The cooling time scale of the interface
between the dense sheath and hot cocoon can be
affected by the thermal conduction. However,
since the width of this layer is thin except
models for low Mach number jets, cooling of the
hot plasma does not change the shape of the
interface in high Mach number jets.

Since the temperature of the shock compressed HI gas does 
not exceed $1000\ \mathrm{K}$, the ionization rate will be negligible during
the formation of the dense, cold sheath from the HI gas.
Inside the jet and cocoon, hydrogen is almost fully ionized 
because $T > 10^{5}\ \mathrm{K}$. Moderately ionized region will appear
in the interface between the hot cocoon and the dense sheath, 
where $T \sim 10^{4}\ \mathrm{K}$. 
This region is formed by the compression of the warm interstellar medium by the bow shock of the jet. 
Since the ionization time scale near the shock front is about $0.1\ \mathrm{Myr}$ and the recombination time scale is about 
$10\ \mathrm{Myr}$, the ionization rate will increase just behind the bow shock , and gradually decrease in the downflow \citep[e.g.,][]{2000ApJ...532..980K}.
Thus, the warm interstellar medium which has been ionized by the bow shock is 
kept partially ionized.

The mass of the jet-aligned molecular clouds estimated from CO observations is $400-2300M_{\odot}$ \citep{2008PASJ...60..715Y}. 
According to the result of our simulation, the total mass of the cold, dense sheath where the number density exceeds $20\ \mathrm{cm^{-3}}$ is $\sim 1200M_{\odot}$ for MC6, which is consistent with observations. 
The cold dense gas in the sheath expands in the radial direction with speed $\sim 1-2\ \mathrm{km\ s^{-1}}$. 
This expansion takes place because the shock heated hot ($T \sim 10^{6}\ \mathrm{K}$) gas in the cocoon (backflow region) pushes the sheath in the radial direction. 
This expansion speed in our simulation is consistent with the speed of the molecular gas ($2-5\ \mathrm{km\ s^{-1}}$) obtained from CO observations \citep{2008PASJ...60..715Y}.

The interface between the sheath and the backflow wiggles because the 
Kelvin-Helmholtz instability grows by the velocity shear between the sheath and the
backflow. 
Since the toroidal magnetic field is accumulated in this boundary, this layer may subject to the interchange instability. 
Although magnetic fields along the jet axis are not amplified in axisymmetric simulations, they can be amplified if the radial components are generated by non-axisymmetric motions and stretched by the velocity shear around the interface. 
We need to carry out three-dimensional MHD simulations to study the magnetic field amplification by these mechanisms and their effects on the structure and stability of the sheath. 
We would like to report the results of 3D MHD simulations in subsequent papers.

We thank T. Hanawa, Y. Matsumoto, S. Miyaji, James M. Stone, M. Machida, A. Mizuta and K. Suzuki for discussion. 
Numerical computations were carried out by FX1 at JAXA and XT4 and XC30 at Center for Computational Astrophysics, NAOJ. 
This work is supported by grants in aid for scientific research by JSPS Core-to-Core No. 22001, JSPS KAKENHI Grant Number (21253003, 23340040) and Grant-in-Aid for JSPS Fellow (24.4786).

\bigskip

\bibliography{apj-jour,letterapj}

\begin{thebibliography}{44}
\expandafter\ifx\csname natexlab\endcsname\relax\def\natexlab#1{#1}\fi

\bibitem[{{Bertsch} {et~al.}(1993){Bertsch}, {Dame}, {Fichtel}, {Hunter},
  {Sreekumar}, {Stacy}, \& {Thaddeus}}]{1993ApJ...416..587B}
{Bertsch}, D.~L., {Dame}, T.~M., {Fichtel}, C.~E., {et~al.} 1993, \apj, 416,
  587

\bibitem[{{Blondin} {et~al.}(1990){Blondin}, {Fryxell}, \&
  {K{\"o}nigl}}]{1990ApJ...360..370B}
{Blondin}, J.~M., {Fryxell}, B.~A., \& {K{\"o}nigl}, A. 1990, \apj, 360, 370

\bibitem[{{Blondin} {et~al.}(1989){Blondin}, {K{\"o}nigl}, \&
  {Fryxell}}]{1989ApJ...337L..37B}
{Blondin}, J.~M., {K{\"o}nigl}, A., \& {Fryxell}, B.~A. 1989, \apjl, 337, L37

\bibitem[{{Brinkmann} {et~al.}(1996){Brinkmann}, {Aschenbach}, \&
  {Kawai}}]{1996A&A...312..306B}
{Brinkmann}, W., {Aschenbach}, B., \& {Kawai}, N. 1996, \aap, 312, 306

\bibitem[{{Brinkmann} {et~al.}(1991){Brinkmann}, {Kawai}, {Matsuoka}, \&
  {Fink}}]{1991A&A...241..112B}
{Brinkmann}, W., {Kawai}, N., {Matsuoka}, M., \& {Fink}, H.~H. 1991, \aap, 241,
  112

\bibitem[{{Brinkmann} {et~al.}(2007){Brinkmann}, {Pratt}, {Rohr}, {Kawai}, \&
  {Burwitz}}]{2007A&A...463..611B}
{Brinkmann}, W., {Pratt}, G.~W., {Rohr}, S., {Kawai}, N., \& {Burwitz}, V.
  2007, \aap, 463, 611

\bibitem[{{Clarke} {et~al.}(1986){Clarke}, {Norman}, \&
  {Burns}}]{1986ApJ...311L..63C}
{Clarke}, D.~A., {Norman}, M.~L., \& {Burns}, J.~O. 1986, \apjl, 311, L63

\bibitem[{{Cseh} {et~al.}(2012){Cseh}, {Corbel}, {Kaaret}, {Lang}, {Gris{\'e}},
  {Paragi}, {Tzioumis}, {Tudose}, \& {Feng}}]{2012ApJ...749...17C}
{Cseh}, D., {Corbel}, S., {Kaaret}, P., {et~al.} 2012, \apj, 749, 17

\bibitem[{{Dedner} {et~al.}(2002){Dedner}, {Kemm}, {Kr{\"o}ner}, {Munz},
  {Schnitzer}, \& {Wesenberg}}]{2002JComp...175...645}
{Dedner}, A., {Kemm}, F., {Kr{\"o}ner}, D., {et~al.} 2002, J. Comp. Phys, 175,
  645

\bibitem[{{Downes} {et~al.}(1981){Downes}, {Salter}, \&
  {Pauls}}]{1981A&A....97..296D}
{Downes}, A.~J.~B., {Salter}, C.~J., \& {Pauls}, T. 1981, \aap, 97, 296

\bibitem[{{Dubner} {et~al.}(1998){Dubner}, {Holdaway}, {Goss}, \&
  {Mirabel}}]{1998AJ....116.1842D}
{Dubner}, G.~M., {Holdaway}, M., {Goss}, W.~M., \& {Mirabel}, I.~F. 1998, \aj,
  116, 1842

\bibitem[{{Elston} \& {Baum}(1987)}]{1987AJ.....94.1633E}
{Elston}, R., \& {Baum}, S. 1987, \aj, 94, 1633

\bibitem[{{Fragile} {et~al.}(2004){Fragile}, {Murray}, {Anninos}, \& {van
  Breugel}}]{2004ApJ...604...74F}
{Fragile}, P.~C., {Murray}, S.~D., {Anninos}, P., \& {van Breugel}, W. 2004,
  \apj, 604, 74

\bibitem[{{Frank} {et~al.}(1998){Frank}, {Ryu}, {Jones}, \&
  {Noriega-Crespo}}]{1998ApJ...494L..79F}
{Frank}, A., {Ryu}, D., {Jones}, T.~W., \& {Noriega-Crespo}, A. 1998, \apjl,
  494, L79

\bibitem[{{Geldzahler} {et~al.}(1980){Geldzahler}, {Pauls}, \&
  {Salter}}]{1980A&A....84..237G}
{Geldzahler}, B.~J., {Pauls}, T., \& {Salter}, C.~J. 1980, \aap, 84, 237

\bibitem[{{Goodall} {et~al.}(2011){Goodall}, {Alouani-Bibi}, \&
  {Blundell}}]{2011MNRAS.414.2838G}
{Goodall}, P.~T., {Alouani-Bibi}, F., \& {Blundell}, K.~M. 2011, \mnras, 414,
  2838

\bibitem[{{Harten} {et~al.}(1983){Harten}, {Lax}, \& {van
  Leer}}]{1983SIAM...25..35H}
{Harten}, A., {Lax}, P.~D., \& {van Leer}, B. 1983, SIAM Rev., 25, 35

\bibitem[{{Inoue} {et~al.}(2006){Inoue}, {Inutsuka}, \&
  {Koyama}}]{2006ApJ...652.1331I}
{Inoue}, T., {Inutsuka}, S.-i., \& {Koyama}, H. 2006, \apj, 652, 1331

\bibitem[{{Inoue} {et~al.}(2009){Inoue}, {Yamazaki}, \&
  {Inutsuka}}]{2009ApJ...695..825I}
{Inoue}, T., {Yamazaki}, R., \& {Inutsuka}, S.-i. 2009, \apj, 695, 825

\bibitem[{{Kawashima} {et~al.}(2009){Kawashima}, {Ohsuga}, {Mineshige},
  {Heinzeller}, {Takabe}, \& {Matsumoto}}]{2009PASJ...61..769K}
{Kawashima}, T., {Ohsuga}, K., {Mineshige}, S., {et~al.} 2009, \pasj, 61, 769

\bibitem[{{K{\"o}ssl} {et~al.}(1990){K{\"o}ssl}, {M{\"u}ller}, \&
  {Hillebrandt}}]{1990A&A...229..378K}
{K{\"o}ssl}, D., {M{\"u}ller}, E., \& {Hillebrandt}, W. 1990, \aap, 229, 378

\bibitem[{{Koyama} \& {Inutsuka}(2000)}]{2000ApJ...532..980K}
{Koyama}, H., \& {Inutsuka}, S.-I. 2000, \apj, 532, 980

\bibitem[{{Lebrun} {et~al.}(1983){Lebrun}, {Bennett}, {Bignami}, {Caraveo},
  {Bloemen}, {Hermsen}, {Buccheri}, {Gottwald}, {Kanbach}, \&
  {Mayer-Hasselwander}}]{1983ApJ...274..231L}
{Lebrun}, F., {Bennett}, K., {Bignami}, G.~F., {et~al.} 1983, \apj, 274, 231

\bibitem[{{Lind} {et~al.}(1989){Lind}, {Payne}, {Meier}, \&
  {Blandford}}]{1989ApJ...344...89L}
{Lind}, K.~R., {Payne}, D.~G., {Meier}, D.~L., \& {Blandford}, R.~D. 1989,
  \apj, 344, 89

\bibitem[{{Margon}(1984)}]{1984ARA&A..22..507M}
{Margon}, B. 1984, \araa, 22, 507

\bibitem[{{Margon} \& {Anderson}(1989)}]{1989ApJ...347..448M}
{Margon}, B., \& {Anderson}, S.~F. 1989, \apj, 347, 448

\bibitem[{{Marshall} {et~al.}(2002){Marshall}, {Canizares}, \&
  {Schulz}}]{2002ApJ...564..941M}
{Marshall}, H.~L., {Canizares}, C.~R., \& {Schulz}, N.~S. 2002, \apj, 564, 941

\bibitem[{{Matsumoto} {et~al.}(2013){Matsumoto}, {Amano}, \&
  {Hoshino}}]{2013PhRvL.111u5003M}
{Matsumoto}, Y., {Amano}, T., \& {Hoshino}, M. 2013, Physical Review Letters,
  111, 215003

\bibitem[{{Miyoshi} \& {Kusano}(2005)}]{2005JComp...208...315}
{Miyoshi}, T., \& {Kusano}, K. 2005, J. Comp. Phys, 208, 315

\bibitem[{{Monceau-Baroux} {et~al.}(2014){Monceau-Baroux}, {Porth}, {Meliani},
  \& {Keppens}}]{2014A&A...561A..30M}
{Monceau-Baroux}, R., {Porth}, O., {Meliani}, Z., \& {Keppens}, R. 2014, \aap,
  561, A30

\bibitem[{{Norman} {et~al.}(1982){Norman}, {Winkler}, {Smarr}, \&
  {Smith}}]{1982A&A...113...285N}
{Norman}, M.~L., {Winkler}, K.-H.~A., {Smarr}, L., \& {Smith}, M.~D. 1982,
  \aap, 113, 285

\bibitem[{{Pakull} {et~al.}(2010){Pakull}, {Soria}, \&
  {Motch}}]{2010Natur.466..209P}
{Pakull}, M.~W., {Soria}, R., \& {Motch}, C. 2010, \nat, 466, 209

\bibitem[{{Richings} {et~al.}(2014){Richings}, {Schaye}, \&
  {Oppenheimer}}]{2014MNRAS.440.3349R}
{Richings}, A.~J., {Schaye}, J., \& {Oppenheimer}, B.~D. 2014, \mnras, 440,
  3349

\bibitem[{{Roe}(1981)}]{1981JCoPh..43..357R}
{Roe}, P.~L. 1981, J. Comp. Phys, 43, 357

\bibitem[{{Safi-Harb} \& {Oegelman}(1997)}]{1997ApJ...483..868S}
{Safi-Harb}, S., \& {Oegelman}, H. 1997, \apj, 483, 868

\bibitem[{{Soria} {et~al.}(2010){Soria}, {Pakull}, {Broderick}, {Corbel}, \&
  {Motch}}]{2010MNRAS.409..541S}
{Soria}, R., {Pakull}, M.~W., {Broderick}, J.~W., {Corbel}, S., \& {Motch}, C.
  2010, \mnras, 409, 541

\bibitem[{{Stone} \& {Hardee}(2000)}]{2000ApJ...540..192S}
{Stone}, J.~M., \& {Hardee}, P.~E. 2000, \apj, 540, 192

\bibitem[{{Stone} \& {Norman}(1993)}]{1993ApJ...413..198S}
{Stone}, J.~M., \& {Norman}, M.~L. 1993, \apj, 413, 198

\bibitem[{{Sutherland} \& {Dopita}(1993)}]{1993ApJS...88..253S}
{Sutherland}, R.~S., \& {Dopita}, M.~A. 1993, \apjs, 88, 253

\bibitem[{{Te{\c s}ileanu} {et~al.}(2008){Te{\c s}ileanu}, {Mignone}, \&
  {Massaglia}}]{2008A&A...488..429T}
{Te{\c s}ileanu}, O., {Mignone}, A., \& {Massaglia}, S. 2008, \aap, 488, 429

\bibitem[{{Todo} {et~al.}(1992){Todo}, {Uchida}, {Sato}, \&
  {Rosner}}]{1992PASJ...44...245T}
{Todo}, Y., {Uchida}, Y., {Sato}, T., \& {Rosner}, R. 1992, \pasj, 44, 245

\bibitem[{{Todo} {et~al.}(1993){Todo}, {Uchida}, {Sato}, \&
  {Rosner}}]{1993ApJ...403...164}
---. 1993, \apj, 403, 164

\bibitem[{{Yamamoto} {et~al.}(2008){Yamamoto}, {Ito}, {Ishigami}, {Fujishita},
  {Kawase}, {Kawamura}, {Mizuno}, {Onishi}, {Mizuno}, {McClure-Griffiths}, \&
  {Fukui}}]{2008PASJ...60..715Y}
{Yamamoto}, H., {Ito}, S., {Ishigami}, S., {et~al.} 2008, \pasj, 60, 715

\bibitem[{{Yamauchi} {et~al.}(1994){Yamauchi}, {Kawai}, \&
  {Aoki}}]{1994PASJ...46L.109Y}
{Yamauchi}, S., {Kawai}, N., \& {Aoki}, T. 1994, \pasj, 46, L109

\end{thebibliography}

\newpage

\begin{table}[!ht]
\begin{center}
  \caption{Model parameters.}
  \begin{tabular}{ccccccc} \tableline \tableline
    Model & cooling & $v_{\mathrm{jet}}$ ($\mathrm{km\ s^{-1}}$) & $n_{\mathrm{jet}}$ ($\mathrm{cm^{-3}}$) & $T_{\mathrm{jet}}$ (K) & $\dot{M}_{\mathrm{jet}} (\mathrm{g\ s^{-1}})$ & resolution \\ \tableline
    MA6 & no & 220 & $1.5 \times 10^{-2}$ & $9.3 \times 10^{4}$ & $1.5 \times 10^{19}$ & $500 \times 1960$ \\
    MC3 & yes & 110 & $1.5 \times 10^{-2}$ & $9.3 \times 10^{4}$ & $7.3 \times 10^{18}$ & $900 \times 3920$ \\
    MC6 & yes & 220 & $1.5 \times 10^{-2}$ & $9.3 \times 10^{4}$ & $1.5 \times 10^{19}$ & $900 \times 3920$ \\
    MC6H & yes & 680 & $1.5 \times 10^{-3}$ & $9.3 \times 10^{5}$ & $4.7 \times 10^{18}$ & $900 \times 3920$ \\
    MC12 & yes & 440 & $1.5 \times 10^{-2}$ & $9.3 \times 10^{4}$ & $2.9 \times 10^{19}$ & $900 \times 3920$ \\
    MC19H & yes & 2200 & $1.5 \times 10^{-3}$ & $9.3 \times 10^5$ & $1.5 \times 10^{19}$ & $900 \times 3920$ \\
    \tableline
  \end{tabular}

\end{center}
\end{table}

 \begin{figure*}[!ht]
  \epsscale{1.0}
   \plotone{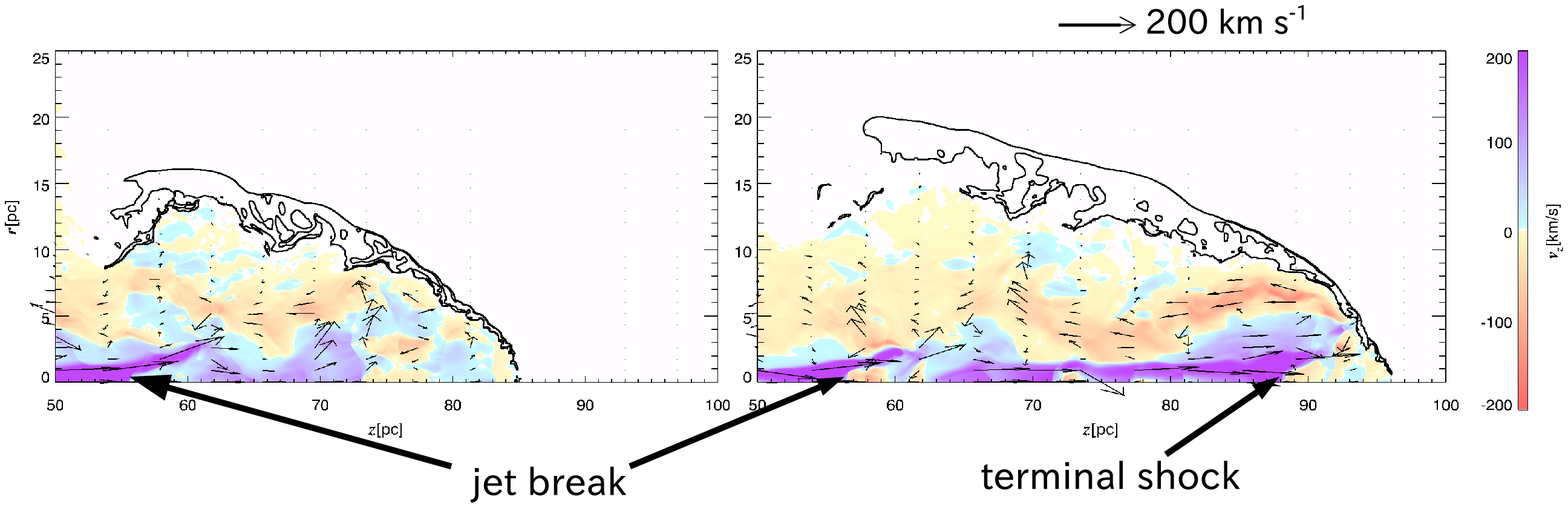}
   \plotone{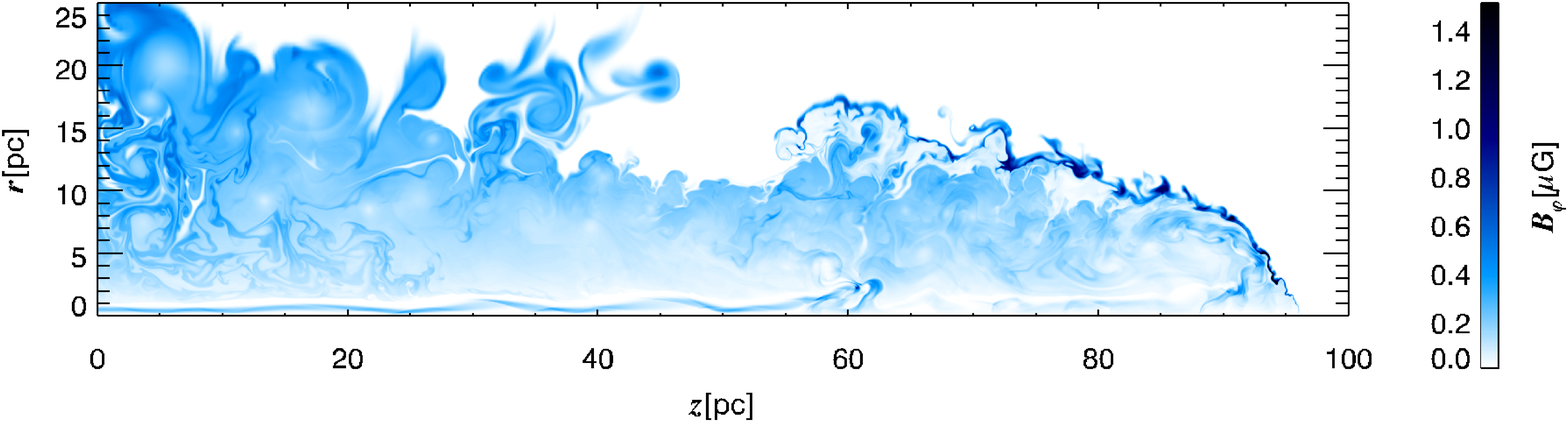}  
   \caption{The top panel shows $v_{z}$ distribution (color) and radial velocity $v_{r} =0.5, 1.0\ \mathrm{km\ s^{-1}}$ of the dense sheath where $n > 10\ \mathrm{cm^{-3}}$ (black contours) and velocity (arrows) for a MHD model with cooling at $10.9\ \mathrm{Myr}$ (top left) and $14.7\ \mathrm{Myr}$ (top right). 
     The bottom panel shows distribution of toroidal magnetic field at $t=14.7\ \mathrm{Myr}$ for MHD simulations including the cooling. \label{jv}}
 \end{figure*}

 \begin{figure*}[!ht]
  \epsscale{1.0}
   \plotone{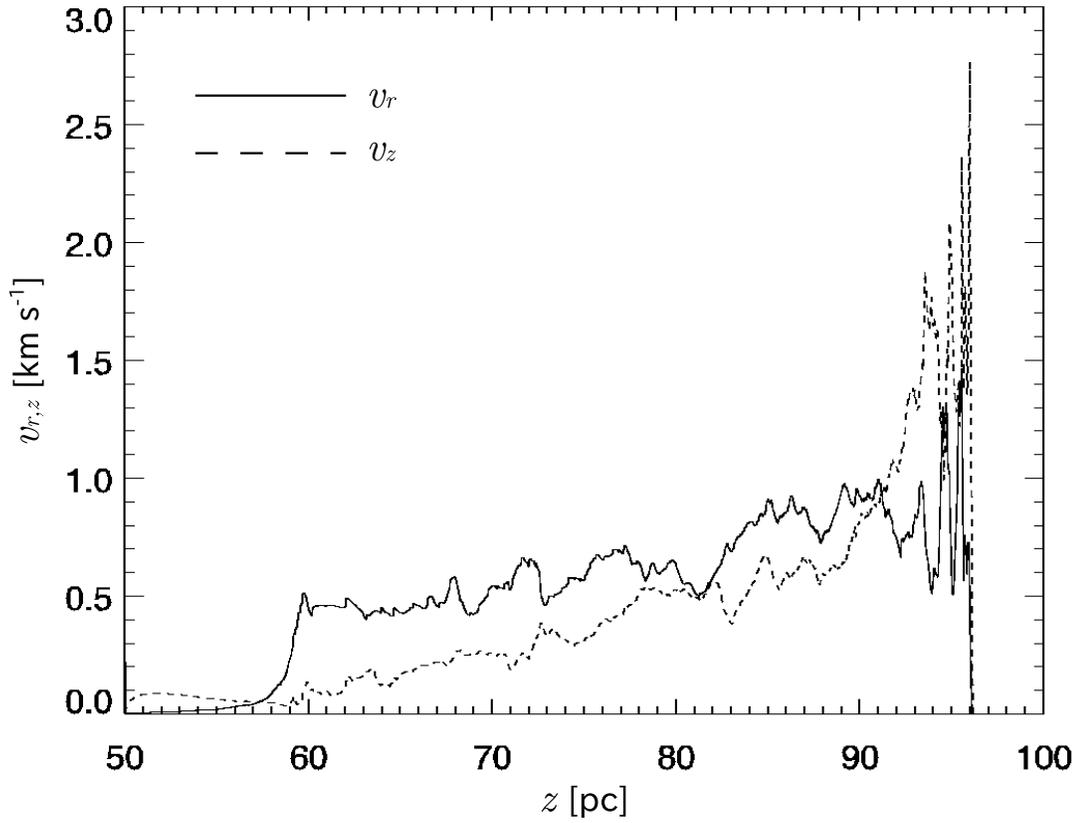}
   \caption{Distribution of mean velocity of high density region ($n>7\ \mathrm{cm^{-3}}$) for MC6. Solid and dashed curves show the radial velocity and the velocity along the jet, respectively. \label{zvrvz}}
 \end{figure*}

 \begin{figure*}[!ht]
  \epsscale{1.0}
   \plotone{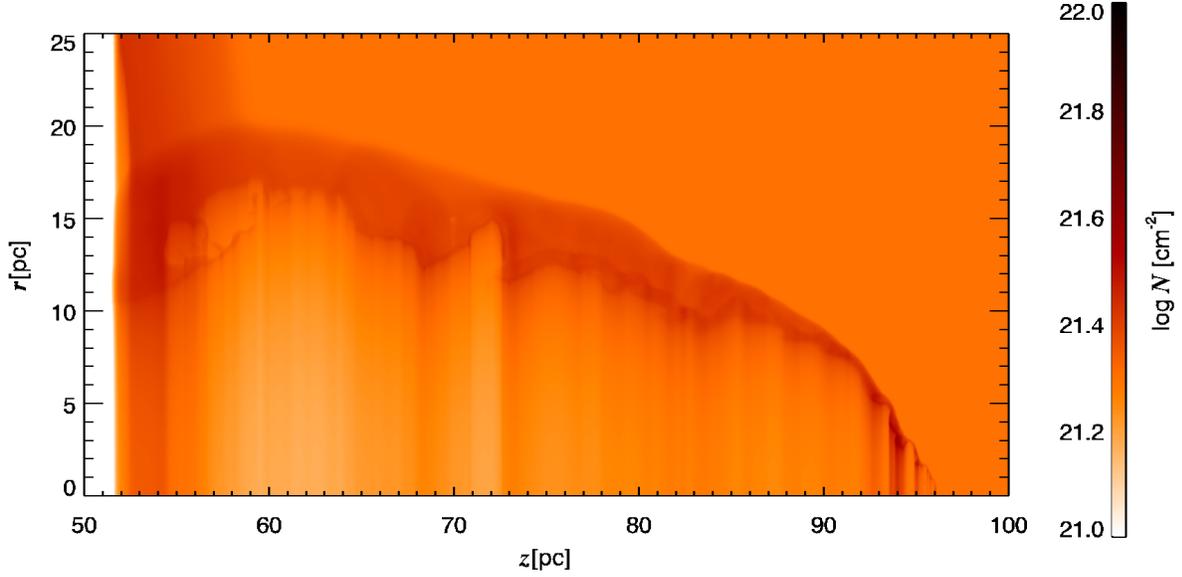}
   \caption{The column number density for MC6 at $t=14.7\ \mathrm{Myr}$. \label{sig}}
 \end{figure*}

 \begin{figure*}[!ht]
  \epsscale{1.0}
   \plotone{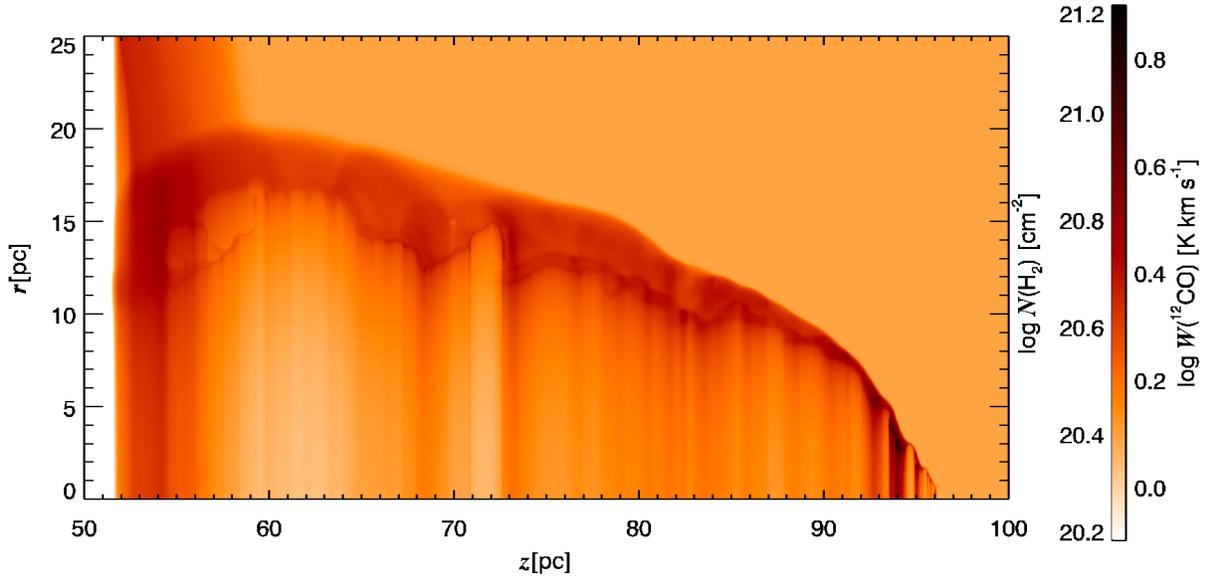}
   \caption{The column number density of $\mathrm{H_{2}}$ and the CO intensity for MC6 at $t=14.7\ \mathrm{Myr}$. \label{sigh}}
 \end{figure*}

 \begin{figure*}[!ht]
  \epsscale{1.0}
   \plotone{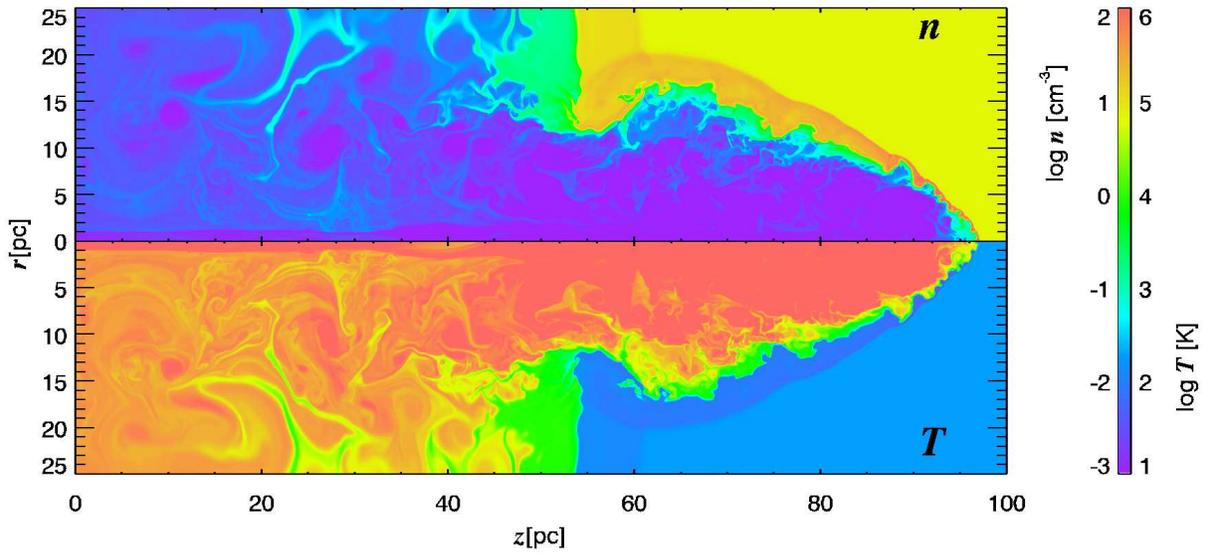}
   \caption{The density and temperature distribution for MC6H at $t=14.7\ \mathrm{Myr}$. \label{jeth}}
 \end{figure*}

\begin{figure*}[!ht]
  \epsscale{1.0}
   \plotone{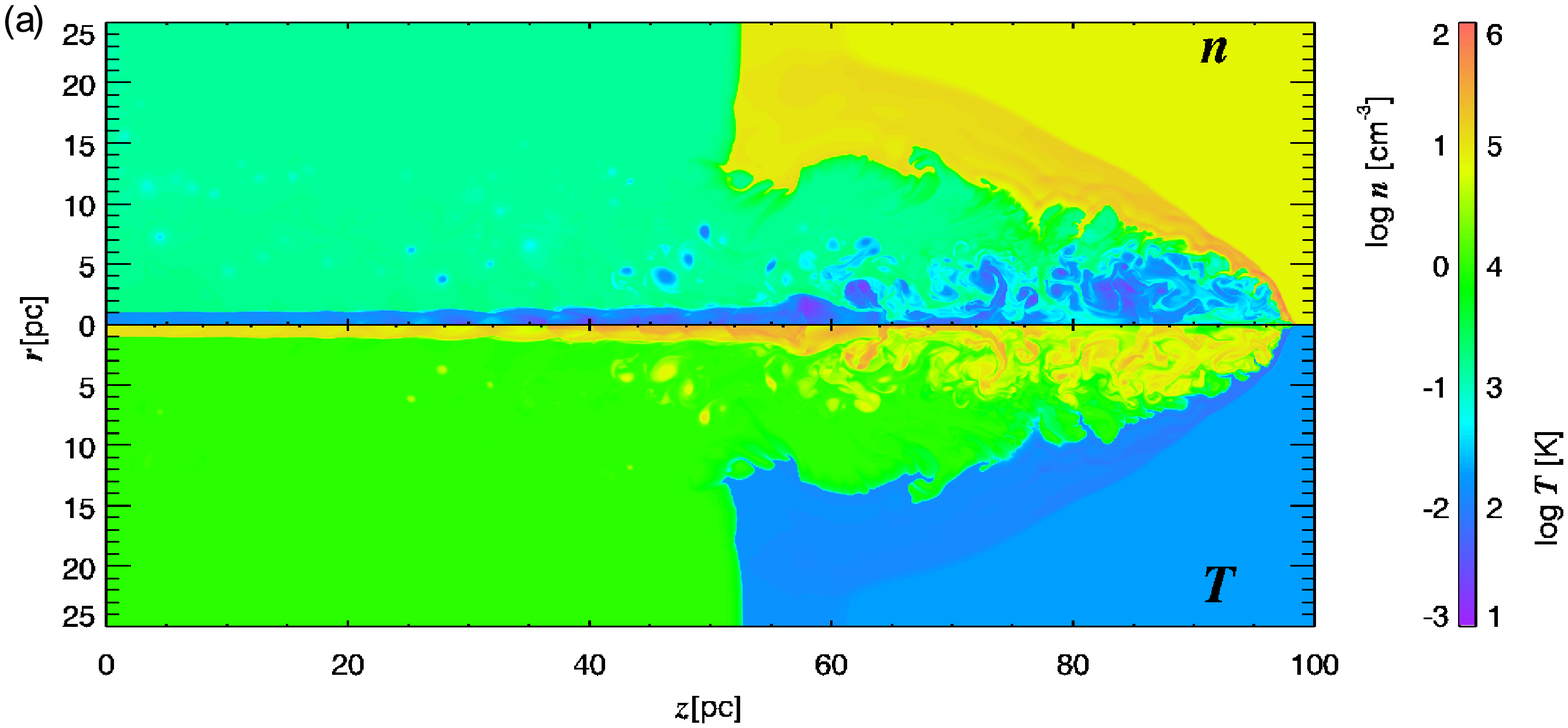} 
   \plotone{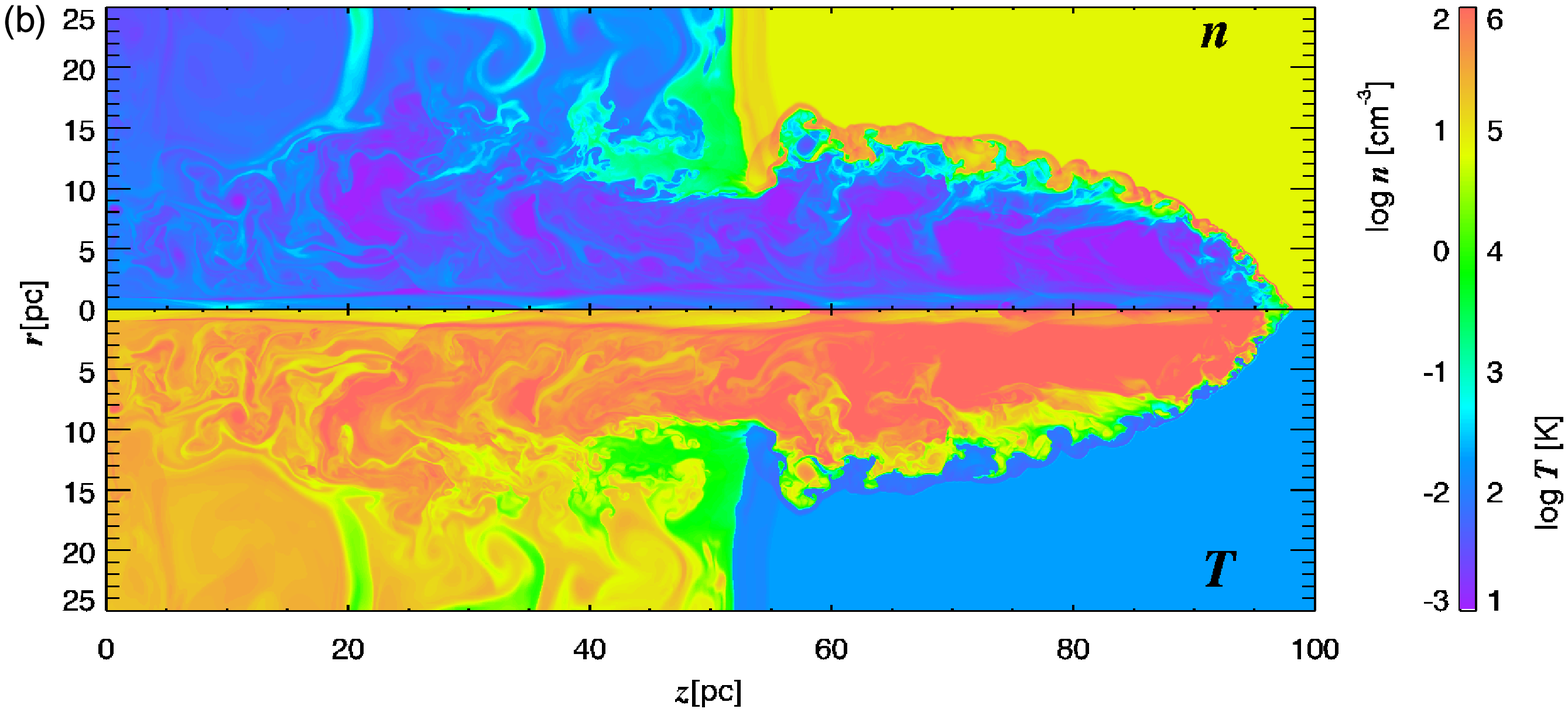}  
  \caption{The density and temperature distribution for 
    (a) MC3 at $t=27.1\ \mathrm{Myr}$ and (b) MC12 at $t=5.84\ \mathrm{Myr}$. \label{MHDc2}}
 \end{figure*}

 \begin{figure*}[!ht]
  \epsscale{1.0}
   \plotone{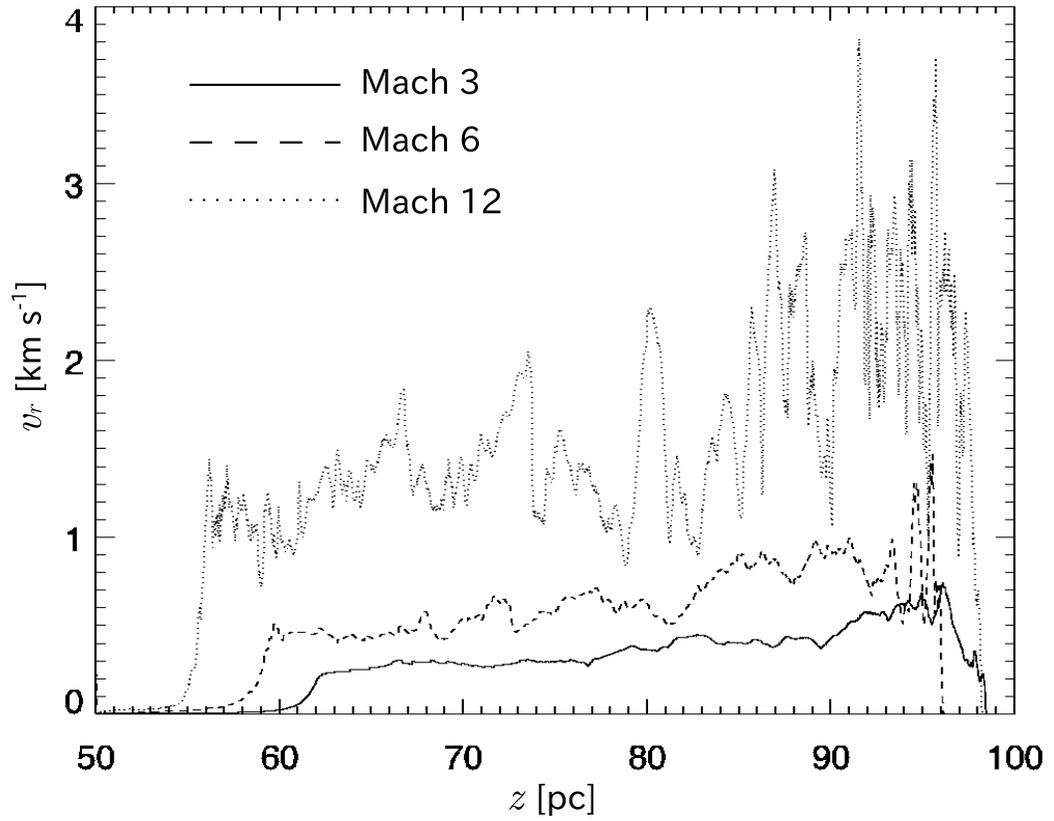}
   \caption{The mean radial velocity of high density region ($n>7\ \mathrm{cm^{-3}}$). Solid, dashed and dotted curves show results for MC3, MC6 and MC12, respectively. \label{zvr}}
 \end{figure*}

 \begin{figure*}[!ht]
  \epsscale{1.0}
   \plotone{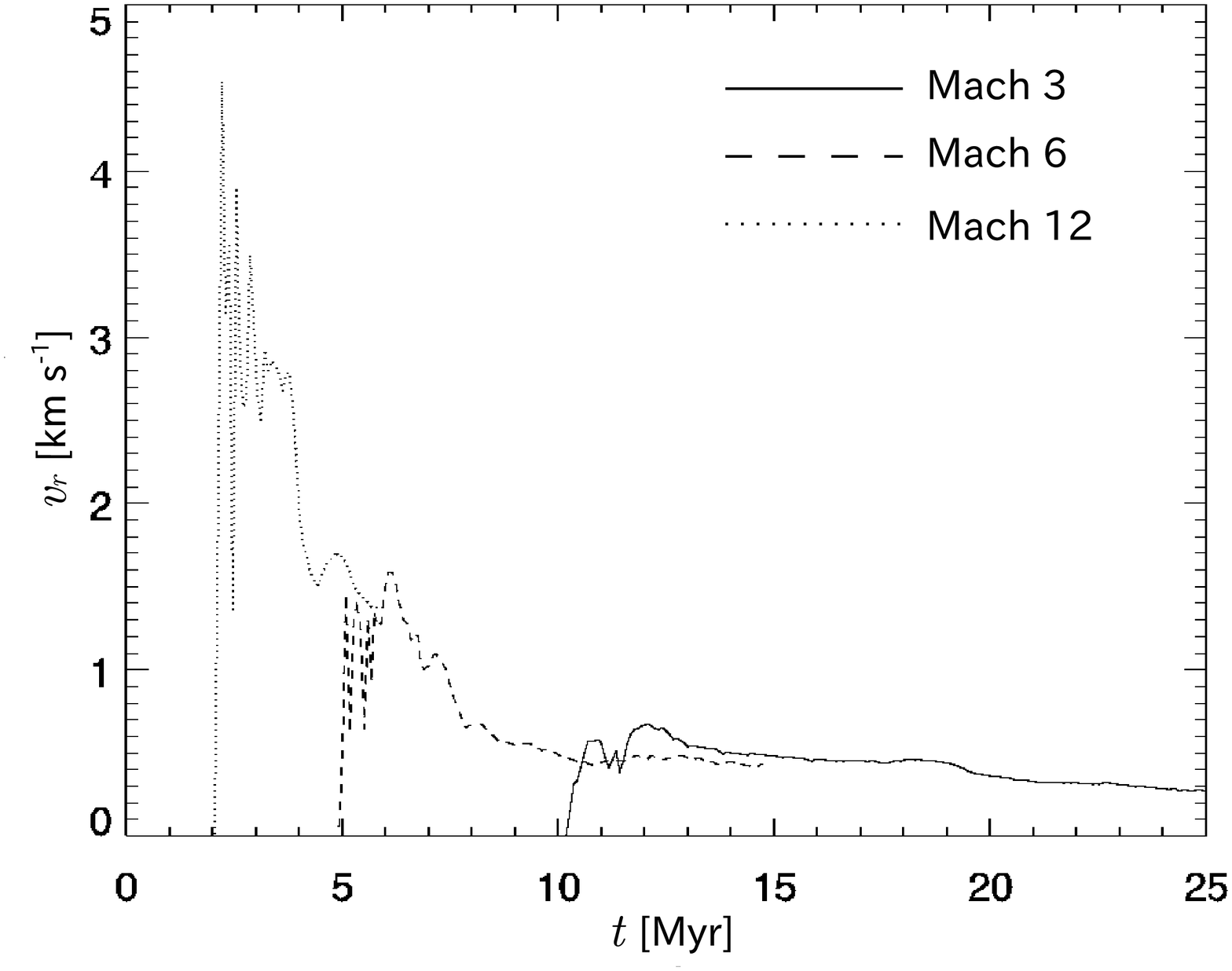}
   \caption{Time evolution of mean radial velocity at $z=65\ \mathrm{pc}$ for model MC3 (solid curve), MC6 (dashed curve) and MC12 (dotted curve). \label{tvr}}
 \end{figure*}

 \begin{figure*}[!ht]
  \epsscale{1.0}
   \plotone{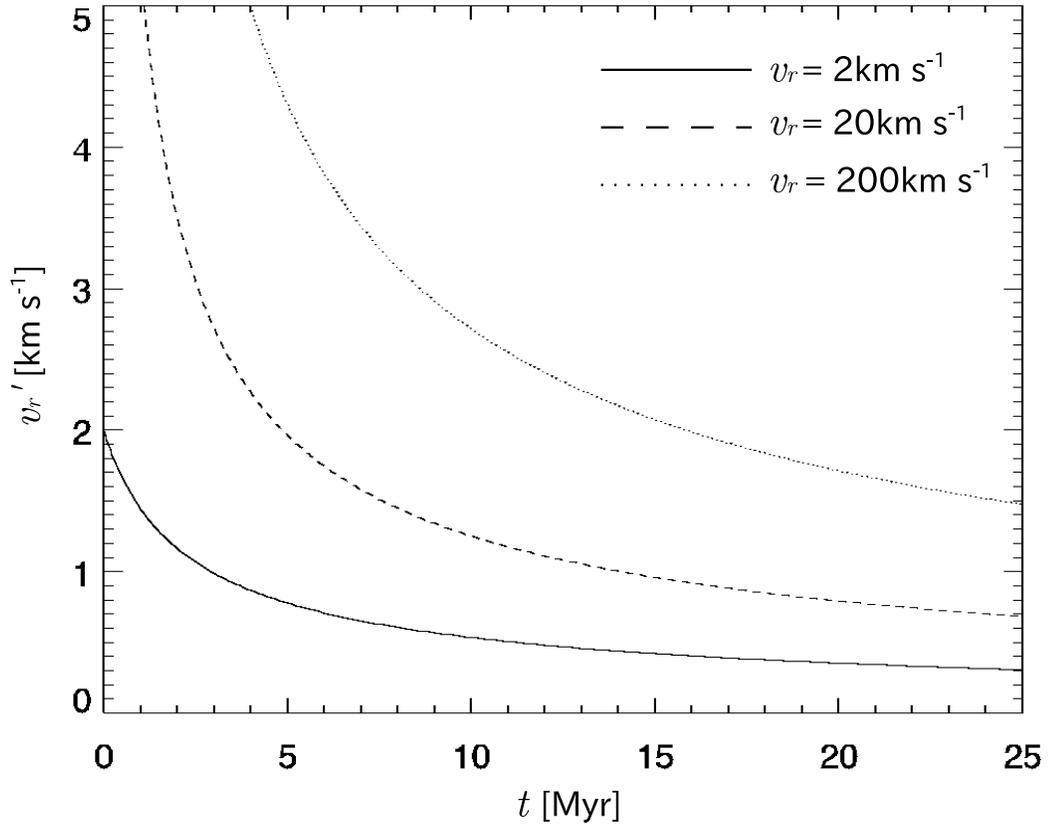}
   \caption{Time evolution of the radial velocity for initial radial velocities $2\ \mathrm{km\ s^{-1}}$ (solid curve), $20\ \mathrm{km\ s^{-1}}$ (dashed curve) and $200\ \mathrm{km\ s^{-1}}$ (dotted curve). \label{vrest}}
 \end{figure*}

 \begin{figure*}[!ht]
  \epsscale{1.0}
   \plotone{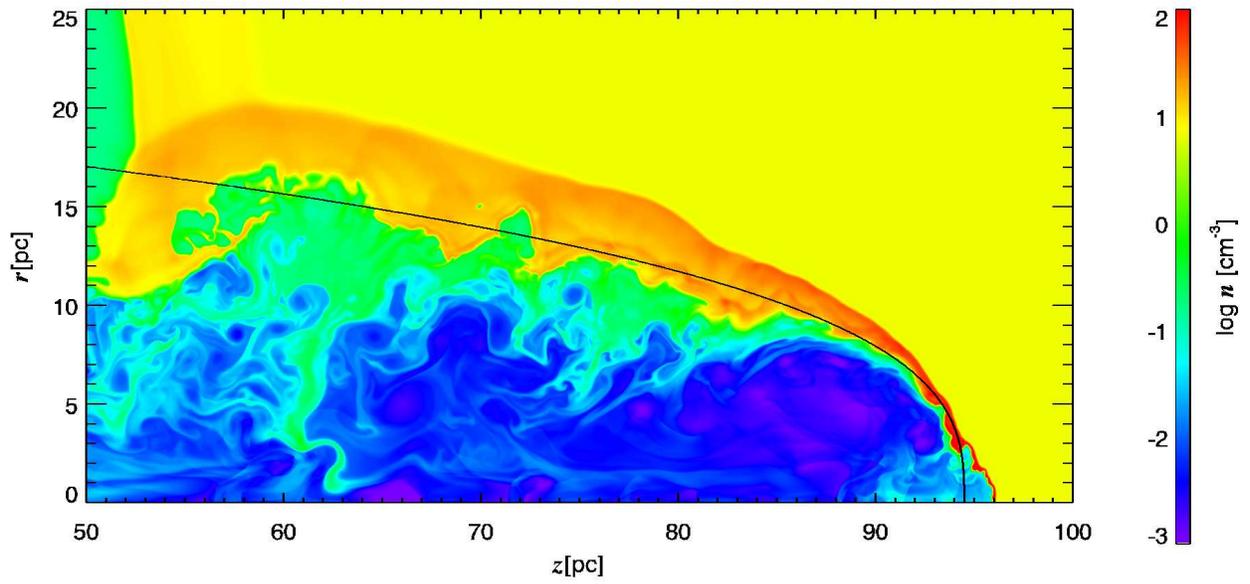}
   \caption{Density distribution for MC6. The black curve shows the location of the interface estimated by equation (23) in the text. \label{intf}}
 \end{figure*}

 \begin{figure*}[!ht]
  \epsscale{1.0}
   \plotone{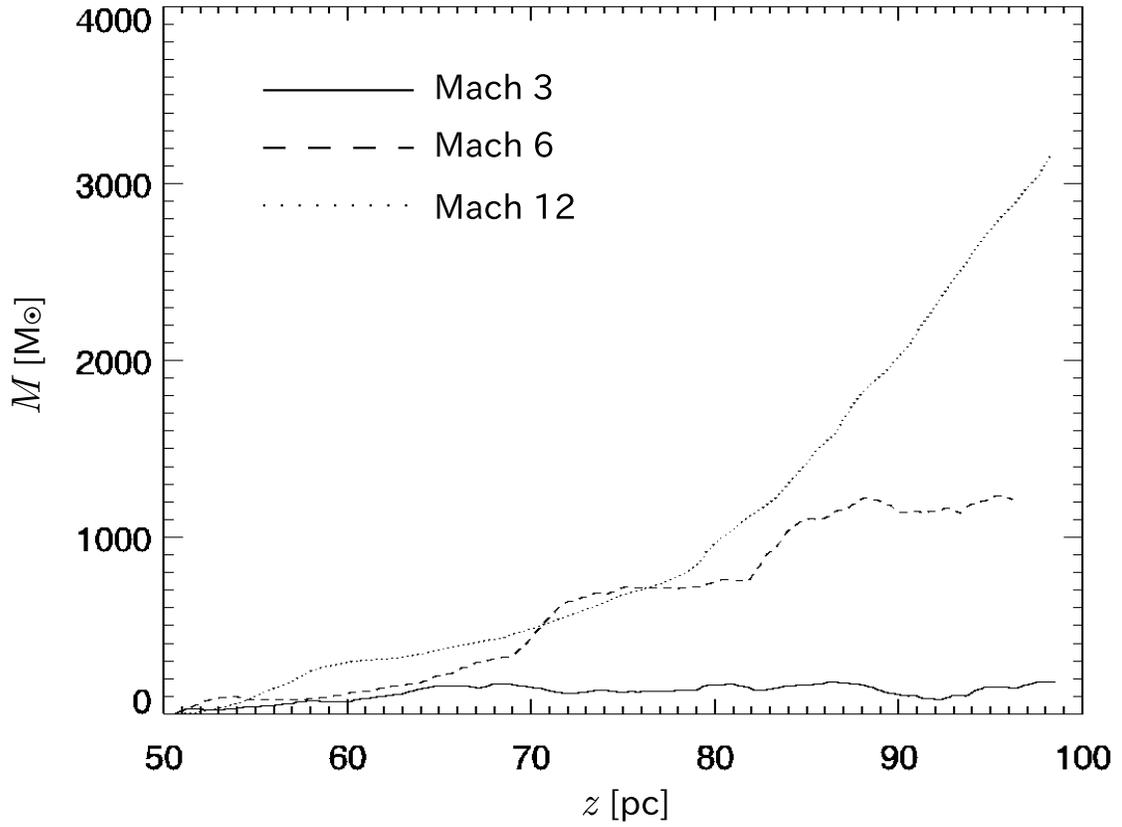}
   \caption{Time evolution of the total mass of the high density gas ($n>20\ \mathrm{cm^{-3}}$). The horizontal axis is the position of the jet head. \label{jtmass}}
 \end{figure*}

 \begin{figure*}[!ht]
  \epsscale{1.0}
   \plotone{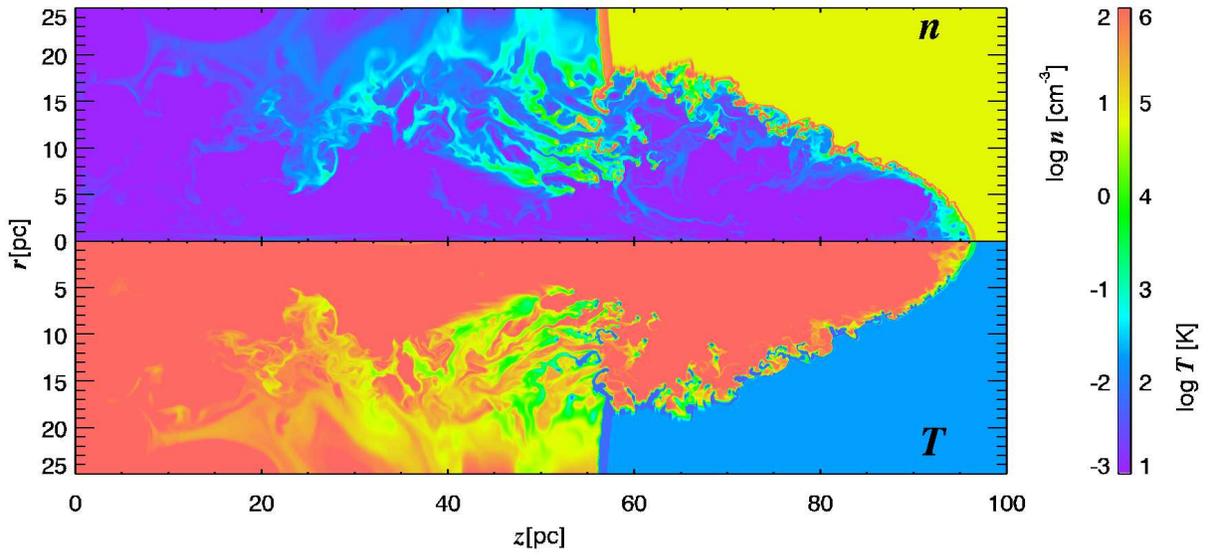}
   \caption{The density and temperature distribution for MC19H at $t=3.53\ \mathrm{Myr}$. \label{jetdmj}}
 \end{figure*}

\end{document}